\documentclass[aps,prl,twocolumn,10pt,superscriptaddress,showpacs]{revtex4-2}
\usepackage{amsmath,amssymb,graphics,epsfig,epstopdf,color,verbatim,ulem,braket,tabularx}
\usepackage{blindtext}
\usepackage[colorlinks,linkcolor=blue,citecolor=blue,urlcolor=blue]{hyperref}

\graphicspath{{Figures/}}

\begin{document}

\title{Condensate Fraction Scaling and Berezinskii-Kosterlitz-Thouless Transition of Superconductivity and Superfluidity}

\author{Yuan-Yao He}
\email{heyuanyao@nwu.edu.cn}
\affiliation{Institute of Modern Physics, Northwest University, Xi'an 710127, China}
\affiliation{Shaanxi Key Laboratory for Theoretical Physics Frontiers, Xi'an 710127, China}
\affiliation{Fundamental Discipline Research Center for Quantum Science and Technology of Shaanxi Province, Xi'an 710127, China}
\affiliation{Hefei National Laboratory, Hefei 230088, China}

\begin{abstract}
Characterizing the superconducting and superfluid transitions in two-dimensional (2D) many-body systems is of broad interest and remains a fundamental issue. In this study, we establish the {\it condensate fraction} scaling as a highly efficient tool to achieve that and accordingly propose efficient schemes to accurately determine the associated Berezinskii-Kosterlitz-Thouless (BKT) transitions. Using the 2D attractive Fermi-Hubbard model as a testbed and applying numerically exact auxiliary-field quantum Monte Carlo simulations, we access unprecedented system sizes (up to $64\times 64 = 4096$ lattice sites) and perform a comprehensive analysis for the temperature dependence and finite-size scaling of {\it condensate fraction} across the BKT transition. We demonstrate that this quantity exhibits algebraic scaling below the transition and exponential scaling above it, with significantly reduced finite-size effects comparing to the extensively studied on-site pairing correlator. This greatly improves the determination of BKT transition using moderate system sizes. We also extract finite-size BKT transition temperature from condensate fraction, and confirm its logarithmic correction on system size. Based on the accurately determined transition, we reveal that the specific heat displays an anomaly, showing a peak at a temperature slightly above BKT transition. Our findings should be generally applicable to 2D fermionic and bosonic systems hosting superconductivity or superfluidity. 
\end{abstract}

\date{\today}
\maketitle

Berezinskii-Kosterlitz-Thouless (BKT) transition~\cite{Berezinsky1972,Kosterlitz1973,Kosterlitz1974} is one of the most intriguing property of two-dimensional (2D) systems with U(1) continuous symmetry and short-range interactions. The transition towards the low-temperature phase is typically characterized~\cite{Ryzhov2017} by the formation of bounded vortex-antivortex pairs and a quasi-long-range order with algebraically decaying correlation function versus distance. The most representative examples are 2D superconductivity and superfluidity, for which the theory also predicts a universal jump of the superfluid density~\cite{Nelson1977} at the BKT transition. These key signatures of BKT physics was experimentally verified in liquid $^4{\rm He}$~\cite{Bishop1978,Ceperley1989} and superconducting films~\cite{Beasley1979,Hebard1980,Wolf1981} soon after the celebrated BKT theory, and more recently in 2D interacting Bose gas~\cite{Hadzibabic2006,Ryu2009,Choi2013} and Fermi gas~\cite{Ries2015,Murthy2015}. The rapid progress in ultracold atom systems has enabled the detailed studies of the BKT transition, and many more experiments with growing precision continue to reveal its other rich properties, such as the phase fluctuations~\cite{Sunami2022}, transport and excitation signatures~\cite{Bohlen2020,Sobirey2021,Julian2021,Sobirey2022}. 

Theoretically, the thorough characterization and precise determination of the BKT transition primarily rely on numerically unbiased calculations~\cite{Arisue2009,Chatelain2014,Ueda2021,Filinov2010,Tomita2002,Tobochnik1979,Ota1992,Ota1995,Gupta1992,Nguyen2021,Wenyu2023,Sandvik2010,Scalettar1989,Moreo1991,Singer1996,Paiva2004,Nakano2006,Paiva2010,Fontenele2022,Xingchuan2023,Wang2023,YuanYao2019,YuanYao2022}. Regarding superconductivity and superfluidity, the 2D attractive Hubbard model serves as the minimal model that captures the BKT physics, as investigated by auxiliary-field quantum Monte Carlo (AFQMC) simulations~\cite{Scalettar1989,Moreo1991,Singer1996,Paiva2004,Nakano2006,Paiva2010,Fontenele2022,Xingchuan2023,Wang2023}. These studies typically estimated the BKT transition temperature through finite-size analyses of the on-site spin-singlet pairing correlation or the superfluid density~\cite{Scalapino1992,*Scalapino1993}. However, both quantities exhibit pronounced finite-size effects, which compromise the reliability of standard scaling approaches. In addition, extracting superfluid density from the dynamical current-current correlation function is computationally demanding and usually yields notably noisy results~\cite{Paiva2004,Paiva2010,Fontenele2022}. All of these studies have been constrained to lattice sizes of up to 400 sites, which are likely insufficient to capture the logarithmic finite-size corrections characteristic of BKT transition~\cite{Tomita2002,Nguyen2021,Ueda2021}. These limitations hinder high-precision determination of BKT transition temperature, even when using numerically exact AFQMC simulations.

Alternatively, condensate fraction is also a fundamental quantity to characterize the superconducting and superfluid states~\cite{Yang1962,Boronat2005,Fukushima2007}, defined as the proportion of Bose-condensed bosons or fermion pairs. It was experimentally measured in liquid $^4{\rm He}$~\cite{Sears1979} system and ultracold Fermi gas~\cite{Zwierlein2004,Inada2008,Mark2012} in three dimensions, and found to behave similarly as the pairing order parameter with its onset identifying the superfluid phase transtion. In these systems, this quantity has also been theoretically computed by mean-field theories~\cite{Giorgini1996,Salasnich2005,Fukushima2007,Anna2011} as well as many-body numerical methods~\cite{Boronat2005,He2020,Jensen2020,Adam2022,Koning2023} at both ground state and finite temperatures. In contrast, it is much less studied for 2D superconductivity and superfluidity in many-body systems especially at finite temperatures. Until now, the theoretical and numerical studies of this quantity mainly concentrate on the ground-state results of 2D interacting Fermi gas~\cite{Luca2007,Shihao2015,Shihao2016}. Then the natural questiones arise as the behaviors of this quantity around the BKT transition and whether it could be used to probe the transition. 

We address the questions by systematically exploring the temperature and size dependences of condensate fraction across the BKT transition in the 2D attractive Hubbard model with large-scale and numerically unbiased AFQMC algorithm. Our calculations reach lattice sizes of up to $4096$ sites, an order of magnitude larger than that in previous studies~\cite{Scalettar1989,Moreo1991,Singer1996,Paiva2004,Nakano2006,Paiva2010,Fontenele2022,Xingchuan2023,Wang2023}. This enables us to perform a finite-size scaling analysis of BKT physics in correlated fermion systems with unprecedented high quality. Our numerical results reveal the algebraic (exponential) scaling of condensate fraction in the superfluid (normal) phase, and establish it as a much more effective probe for characterizing the BKT transition, compared to the long-studied on-site pairing correlator. Then we have computed accurate BKT transition temperature based on the scaling behaviors of condensate fraction. A clear
evidence of specific heat anomaly above the BKT transition is also identified in our calculations.

We study the 2D attractive Fermi-Hubbard model on square lattice with the Hamiltonian as
\begin{equation}\begin{aligned}
\label{eq:ModelHmt}
\hat{H} = 
\sum_{\mathbf{k}\sigma} (\varepsilon_{\mathbf{k}}+\mu) c_{\mathbf{k}\sigma}^+c_{\mathbf{k}\sigma}
+ U\sum_{\mathbf{i}}\Big( \hat{n}_{\mathbf{i}\uparrow}\hat{n}_{\mathbf{i}\downarrow} - \frac{\hat{n}_{\mathbf{i}\uparrow} + \hat{n}_{\mathbf{i}\downarrow}}{2} \Big), \nonumber
\end{aligned}\end{equation}
with $\hat{n}_{\mathbf{i}\sigma}=c_{\mathbf{i}\sigma}^+ c_{\mathbf{i}\sigma}$ as the density operator and $\sigma$ ($=\uparrow$ or $\downarrow$) denoting spin. We only consider the nearest-neighbor hopping $t$ with the energy dispersion $\varepsilon_{\mathbf{k}}=-2t(\cos k_x+\cos k_y)$, where $k_x,k_y$ are the momentum defined in units of $2\pi/L$ with linear system size $L$. The fermion filling is $n=N/N_s$ with $N_s=L^2$ and $N$ as the number of lattice sites and fermions. We set $t$ as the energy unit, and focus on attractive interaction $U<0$ with hole doping $\mu>0$. It is well-known that this model has a superconducting ground state~\cite{Yang1989,Mark2020,Hille2020} and exhibits BKT transition at finite temperatures~\cite{Scalettar1989,Moreo1991,Singer1996,Paiva2004}. We employ state-of-art AFQMC algorithm~\cite{Blankenbecler1981,Hirsch1983,White1989,Scalettar1991,McDaniel2017,Yuanyao2019b,Song2025B} with most recent developments~\cite{Suppl} to study the BKT physics in this model. Our simulations cover a large range of system sizes, from $L=8$ to $L=64$. 

We mainly focus on the condensate fraction and on-site pairing correlator, which both characterize the fermion pairing. The first is initially defined from the macroscopic eigenvalue of the two-body density matrix, expressed as $\boldsymbol{\rho}(\mathbf{r},\mathbf{r}^{\prime})=\{\langle c_{\mathbf{i}\uparrow}^+c_{\mathbf{j}\downarrow}^+c_{\mathbf{i}+\mathbf{r}\uparrow}c_{\mathbf{j}+\mathbf{r}^{\prime}\downarrow}\rangle\}$, in asymptotic limit $|\mathbf{r}|,|\mathbf{r}^{\prime}|\to\infty$~\cite{Yang1962,Boronat2005}. Regarding attractive interaction, we only need to focus on the dominant component of $\boldsymbol{\rho}(\mathbf{r},\mathbf{r}^{\prime})$, i.e., the pairing matrix in momentum space~\cite{Shihao2015,YuanYao2019,YuanYao2022} as $M_{\mathbf{k}\mathbf{k}^{\prime}}=\langle\hat{\Delta}_{\mathbf{k}}^+\hat{\Delta}_{\mathbf{k}^{\prime}}\rangle-\delta_{\mathbf{k}\mathbf{k}^{\prime}}\langle c_{\mathbf{k}\uparrow}^+c_{\mathbf{k}\uparrow}\rangle\langle c_{-\mathbf{k}\downarrow}^+c_{-\mathbf{k \downarrow}}\rangle$, with $\hat{\Delta}_{\mathbf{k}}^+=c_{\mathbf{k}\uparrow}^+c_{-\mathbf{k}\downarrow}^+$ as the $\mathbf{k}$-space spin-singlet pairing operator. Then condensate fraction can be calculated as $n_c=\lambda_{\rm max}/(N/2)$, with $\lambda_{\rm max}$ as 
the leading eigenvalue of $M_{\mathbf{k}\mathbf{k}^{\prime}}$ matrix. The corresponding eigenvector of $\lambda_{\rm max}$ contains the pairing structure information in $\mathbf{k}$ space~\cite{Suppl}. On the other hand, the pairing correlation function is defined as $P(\mathbf{r})=L^{-2}\sum_{\mathbf{i}}\langle\hat{\Delta}_{\mathbf{i}}^+\hat{\Delta}_{\mathbf{i}+\mathbf{r}}+\hat{\Delta}_{\mathbf{i}}\hat{\Delta}_{\mathbf{i}+\mathbf{r}}^+\rangle/4$ with the on-site pairing operator $\hat{\Delta}_{\mathbf{i}}^+=c_{\mathbf{i}\uparrow}^+c_{\mathbf{i}\downarrow}^+$. The corresponding pairing correlator reads $\langle\Delta^2\rangle=L^{-2}\sum_{\mathbf{r}}P(\mathbf{r})$, which has been studied for attractive fermions over decades~\cite{Scalettar1989,Moreo1991,Singer1996,Paiva2004,Nakano2006,Paiva2010,Fontenele2022,Xingchuan2023,Wang2023}. For 2D superfluid phase, $P(\mathbf{r})$ decays algebraically as $P(\mathbf{r})\propto r^{-\eta}$ with $\eta\le\eta_c$ and $\eta_c=1/4$ at BKT transition~\cite{Berezinsky1972,Kosterlitz1973,Kosterlitz1974}, leading to the scaling behavior of $\langle\Delta^2\rangle \propto L^{-\eta}$ in large $L$ limit~\cite{Suppl}. In contrast, they simply decay exponentially in the normal phase.

\begin{figure}[t]
\centering
\includegraphics[width=0.912\columnwidth]{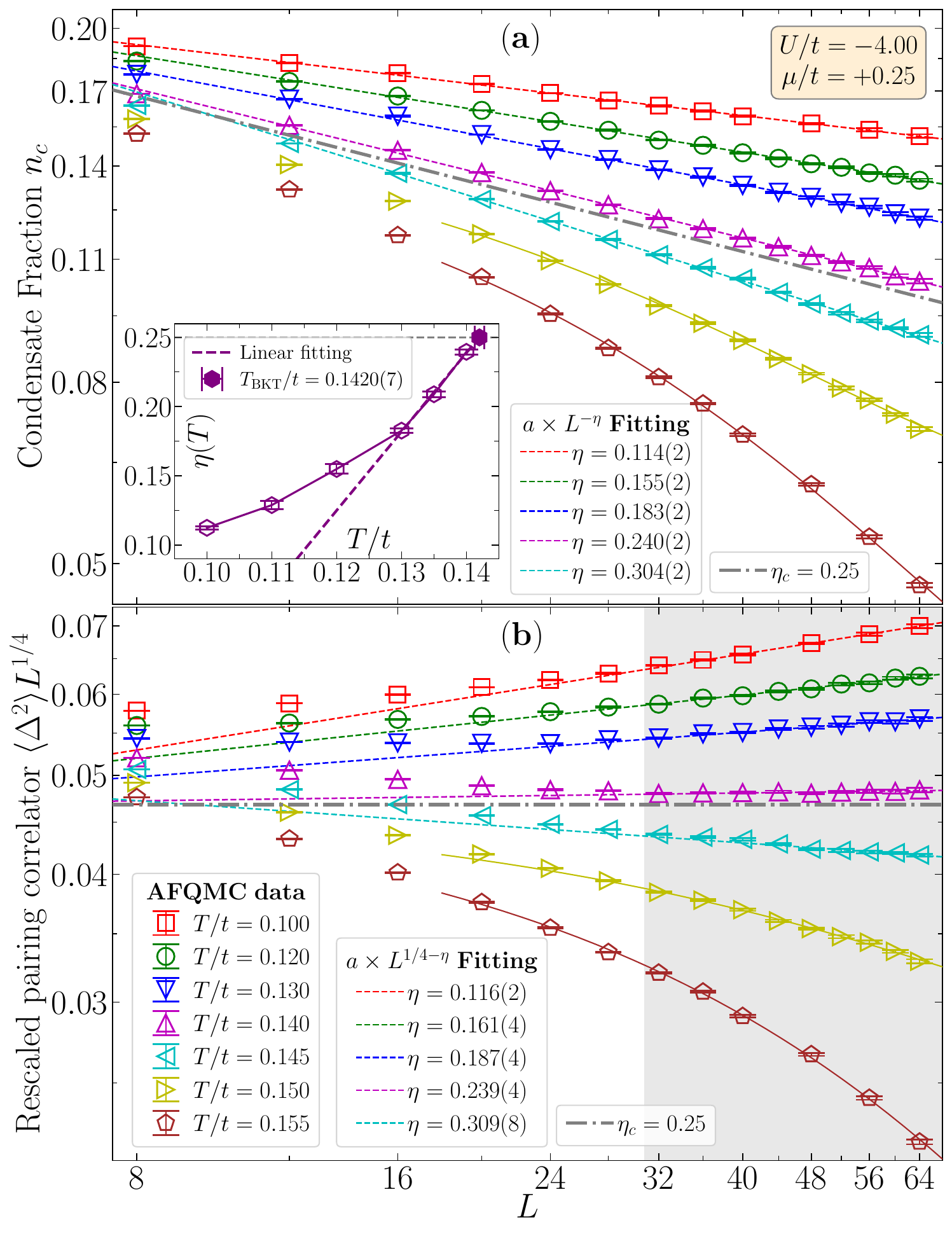}
\caption{The finite-size scaling and log-log plots of (a) condensate fraction $n_c$, and (b) rescaled pairing correlator $\langle\Delta^2\rangle L^{1/4}$, versus linear system size $L$, for the 2D attractive Fermi-Hubbard model with $U/t=-4,\mu/t=0.25$. The AFQMC results in both panels share the same legends. The dashed lines represent power-law fits to the data for $L\ge20$ in panel (a), and with $L\ge32$ in panel (b), while the solid lines denote exponential fits. Consistent values of the exponent $\eta$ are extracted from both quantites for $T/t\le0.145$. The gray, dash-dot lines highlight $\eta_c=1/4$ at the BKT transition, with the transition temperature bounded as $0.140<T_{\rm BKT}/t<0.145$ (with fermion filling $n\simeq0.8003$~\cite{Suppl}). The inset in (a) shows a linear extrapolation of $\eta$ to $\eta_c$ versus $T$, yielding a transition temperature $T_{\rm BKT}/t=0.1420(7)$. The light gray shaded region in (b) marks the scaling regime ($L\ge 32$) of $\langle\Delta^2\rangle$, in which it exhibits the correct size-scaling behavior.}
\label{fig:Fig01CondScaling}
\end{figure}

As the main results of this work, Fig.~\ref{fig:Fig01CondScaling} summarizes finite-size scaling behaviors of condensate fraction $n_c$ and on-site pairing correlator $\langle\Delta^2\rangle$ for $U/t=-4,\mu/t=0.25$. Upon cooling, $n_c$ monotonically grows as evolving from the normal phase to the superfluid phase~\cite{Suppl}. Then the log-log plot and fittings for the numerical results in Fig.~\ref{fig:Fig01CondScaling}(a) clearly reveal that $n_c$ also has algebraic ($\propto L^{-\eta}$) and exponential ($\propto e^{-\alpha L}$) scalings versus $L$ in the low and high temperature regimes, respectively. As a comparison, Fig.~\ref{fig:Fig01CondScaling}(b) plots $\langle\Delta^2\rangle L^{1/4}$ versus $L$, and the results with $L\ge 32$ confirm the size-scaling behaviors of $\langle\Delta^2\rangle$ from BKT theory. As it is straightforward, the horizontal line (gray, dash-dot) denoting $\eta_c=1/4$ in Fig.~\ref{fig:Fig01CondScaling}(b) determines the BKT transition temperature in the range of $0.140$$\sim$$0.145$, which perfectly matches to that from $n_c$ results in Fig.~\ref{fig:Fig01CondScaling}(a). Moreover, the fittings of these two quantities presents well consistent exponents $\eta(T)$ within statistical uncertainty, indicating exactly the same size-scaling behavior of them (in large $L$ limit). Note that at $T/t=0.145$, the algebraic scaling should gradually cross over to exponential decay at larger system sizes even beyond the reach of our AFQMC calculations. These results explicitly illustrate the vanishing $n_c$ and $\langle\Delta^2\rangle$ approaching the thermodynamic limit (TDL) at finite temperatures, thereby confirming the quasi-condensate and quasi-long-range order state in the 2D superfluid phase. Furthermore, the exponent $\eta(T)$ exhibits a linear dependence on $T$ as $T\to T_{\rm BKT}^{-}$~\cite{Nelson1977}. We accordingly perform a linear extrapolation for $\eta(T)$ to the critical $\eta_c$, and reach the transition temperature $T_{\rm BKT}/t=0.1420(7)$ at TDL, as shown in the inset of Fig.~\ref{fig:Fig01CondScaling}(a). 

It is notable that Fig.~\ref{fig:Fig01CondScaling} reveals a prominent discrepancy between $n_c$ and $\langle\Delta^2\rangle$. The former already reaches the scaling regime for $L\ge20$, whereas the latter exhibits the correct asymptotic behavior only for $L\ge32$. $\langle\Delta^2\rangle$ shows pronounced finite-size effects, with results for $L<32$ clearly deviating from the asymptotic scaling observed in larger systems, as shown in Fig.~\ref{fig:Fig01CondScaling}(b). If the numerical simulations are restricted to $L=20$, as in previous studies~\cite{Scalettar1989,Moreo1991,Paiva2004,Nakano2006,Paiva2010,Fontenele2022,Xingchuan2023,Wang2023}, one would obtain the BKT transition temperature in the range $0.12<T_{\rm BKT}/t<0.13$, consistent with that reported in Ref.~\onlinecite{Wang2023}, but with an error exceeding $10\%$ relative to the correct value. Nevertheless, condensate fraction $n_c$ shown in Fig.~\ref{fig:Fig01CondScaling}(a) exhibits much better size-scaling behavior around BKT transition. Results within $L\le20$ can determine the BKT transition temperature, with a small interval of $0.005t$, pinning it down to the quantitatively correct range. These results demonstrate that the condensate fraction exhibits significantly weaker finite-size effects, reaching the asymptotic scaling regime at much smaller system sizes than the on-site pairing correlator. This distinction is crucial for practical numerical studies, where accessible system sizes are limited, as the condensate fraction converges much more rapidly to the asymptotic behavior.

The above scaling difference can be attributed to the distinct underlying physics of these two quantities. Formally, $\langle\Delta^2\rangle$ includes a subleading correction $\propto L^{-2}$~\cite{Suppl}, in addition to the leading term $\propto L^{-\eta}$. Physically, while $\langle\Delta^2\rangle$ only accounts for on-site pairing, the pairing matrix $M_{\mathbf{k}\mathbf{k}^{\prime}}$ originates from $\mathbf{k}$-space pairing, meaning that $n_c$ encompasses contributions from Cooper pairs of all possible sizes. For weak to intermediate attractive interactions, nonlocal fermion pairing is non-negligible, as evidenced by AFQMC results of real-space pair wave function~\cite{Suppl}, and has also been observed in a recent optical lattice experiment~\cite{Hartke2023}. The contribution from local Cooper pairs to the overall pairing can fluctuate significantly with system size. Consequently, the quantity $\langle\Delta^2\rangle$ fails to capture the full pairing structure of the system and it is subject to strong finite-size effects regarding the asymptotic scaling (i.e., $\langle\Delta^2\rangle\propto L^{-\eta}$).

We have obtained similar results as Fig.~\ref{fig:Fig01CondScaling} for different $U/t$ and $\mu/t$, and have verified the same scaling behaviors in fixed-$\mu$ and fixed-density calculations~\cite{Suppl}. We present fixed-$\mu$ results to enable efficient access to large system sizes within the grand-canonical framework, since achieving fixed density requires iterative tuning of $\mu$, which becomes increasingly costly for large $L$. This consideration is particularly relevant here. The on-site pairing correlator has strong finite-size effects and needs $L\ge 32$ to reach the correct scaling regime, making fixed-density calculations much more demanding. In contrast, the condensate fraction yields reliable estimates already at moderate system sizes within the same framework.

\begin{figure}[t]
\centering
\includegraphics[width=0.895\columnwidth]{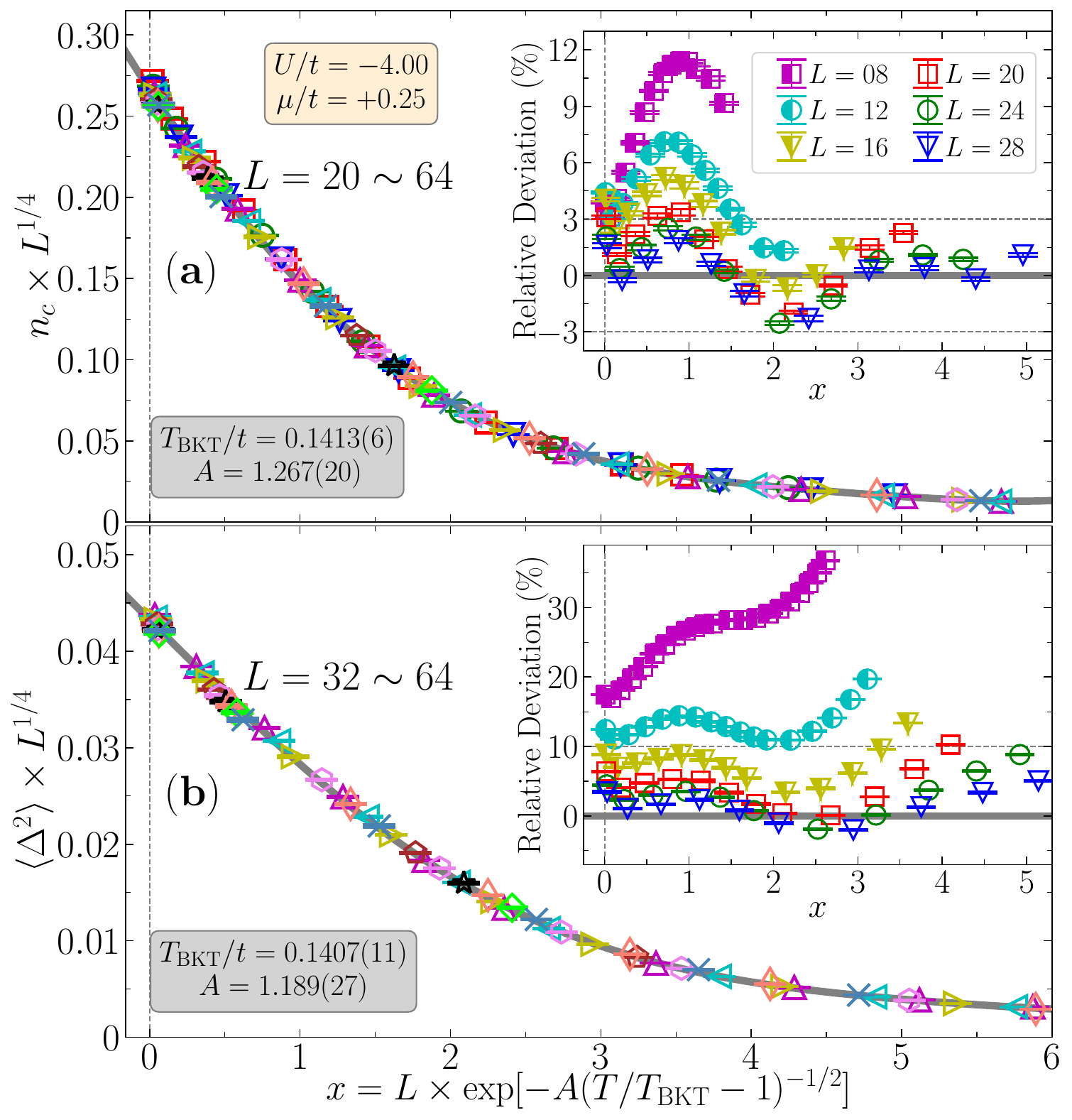}
\caption{Data collapse for (a) condensate fraction $n_c$, and (b) on-site pairing correlator $\langle\Delta^2\rangle$ with fixed exponent $\eta_c=1/4$. The fitting process involves AFQMC data of $L=20$$\sim$$64$ for $n_c$ and $L=32$$\sim$$64$ for $\langle\Delta^2\rangle$, respectively. The gray solid lines are the scaling invariant function $f(x)$, for which we adopt polynomials in $x$ for the fitting. The two insets sharing the same legends plot the relative deviations from $f(x)$ for $L\le28$ results. Simulation parameters are the same as Fig.~\ref{fig:Fig01CondScaling}. }
\label{fig:Fig02Collapse} 
\end{figure}

Our numerical results show that, the finite-size effect of the on-site pairing correlator $\langle\Delta^2\rangle$, manifested as deviation from the correct scaling in small sizes, gets more severe at lower filling and weaker interaction. This can be understood from the fact that the nonlocal pairing becomes more significant in these systems. Instead, such effect is almost absent in condensate fraction $n_c$ with its numerical results always exhibiting nice size-scaling behavior for $L\ge12$ as those shown in Fig.~\ref{fig:Fig01CondScaling}(a). This indicates that accurately determining the BKT transition in 2D interacting Fermi gas~\cite{YuanYao2022} via finite-size analysis of $\langle\Delta^2\rangle$ is considerably more challenging than using $n_c$. All these results suggest that condensate fraction is a more appropriate quantity for characterizing the fermion pairing and determining BKT transition for attracive fermions. Our calculations of superfluid density~\cite{Scalapino1992,Scalapino1993} up to $L=32$ also agree with the BKT transition temperature obtained from the finite-size scaling of the condensate fraction~\cite{Suppl}. 

With the above bounded range, the precise $T_{\rm BKT}$ can be further extracted from standard finite-size scaling. The most commonly used method is via the data collapse of the pairing correlator $\langle\Delta^2\rangle$~\cite{Scalettar1989,Moreo1991,Paiva2004,Nakano2006,Paiva2010,Fontenele2022,Xingchuan2023,Wang2023}. It is based on the scaling relation $\langle\Delta^2\rangle L^{\eta_c}=f(x)$ with $x=L\times\exp[-A(T/T_{\rm BKT}-1)^{-1/2}]$ slightly above the transition as $T\to T_{\rm BKT}^+$~\cite{Paiva2004,Nguyen2021,Wang2023}. The fitting process determines $A$, $\eta_c$ and $T_{\rm BKT}$, which then make the plot of $\langle\Delta^2\rangle L^{\eta_c}$ collapse into a universal curve of $f(x)$. Similarly, condensate fraction $n_c$ should also satisfy the above scaling relation since it has the same scaling behavior as $\langle\Delta^2\rangle$. We then perform data collapse for both $n_c$ and $\langle\Delta^2\rangle$ with fixed $\eta_c=1/4$ for the results presented in Fig.~\ref{fig:Fig01CondScaling}. Considering the finite-size effect in $\langle\Delta^2\rangle$, we use numerical results of $L=32$$\sim$$64$ for $\langle\Delta^2\rangle$ and $L=20$$\sim$$64$ for $n_c$ in the fitting, for which we adopt polynomials for $f(x)$ around $x\to0^+$ and apply the least-squares criterion. The data collapse results are plotted in Fig.~\ref{fig:Fig02Collapse}. Consistent results of $T_{\rm BKT}/t=0.1413(6)$ and $0.1407(11)$ are obtained respectively from $n_c$ and $\langle\Delta^2\rangle$, which agree with the bounded range obtained in Fig.~\ref{fig:Fig01CondScaling}. The insets of Fig.~\ref{fig:Fig02Collapse}, showing relative deviations from $f(x)$ of $L\le28$ results, reveal the more severe finite-size effect in $\langle\Delta^2\rangle$ than that of $n_c$. Specifically, the maximum deviation for $L=20$ is $\sim$$10\%$ in $\langle\Delta^2\rangle$, whereas it is $\sim$$3\%$ for $n_c$. Further including $L\le28$ data in the fitting for $\langle\Delta^2\rangle$ indeed leads to a lower $T_{\rm BKT}$ out of the bounded range. 

\begin{figure}
\centering
\includegraphics[width=0.920\columnwidth]{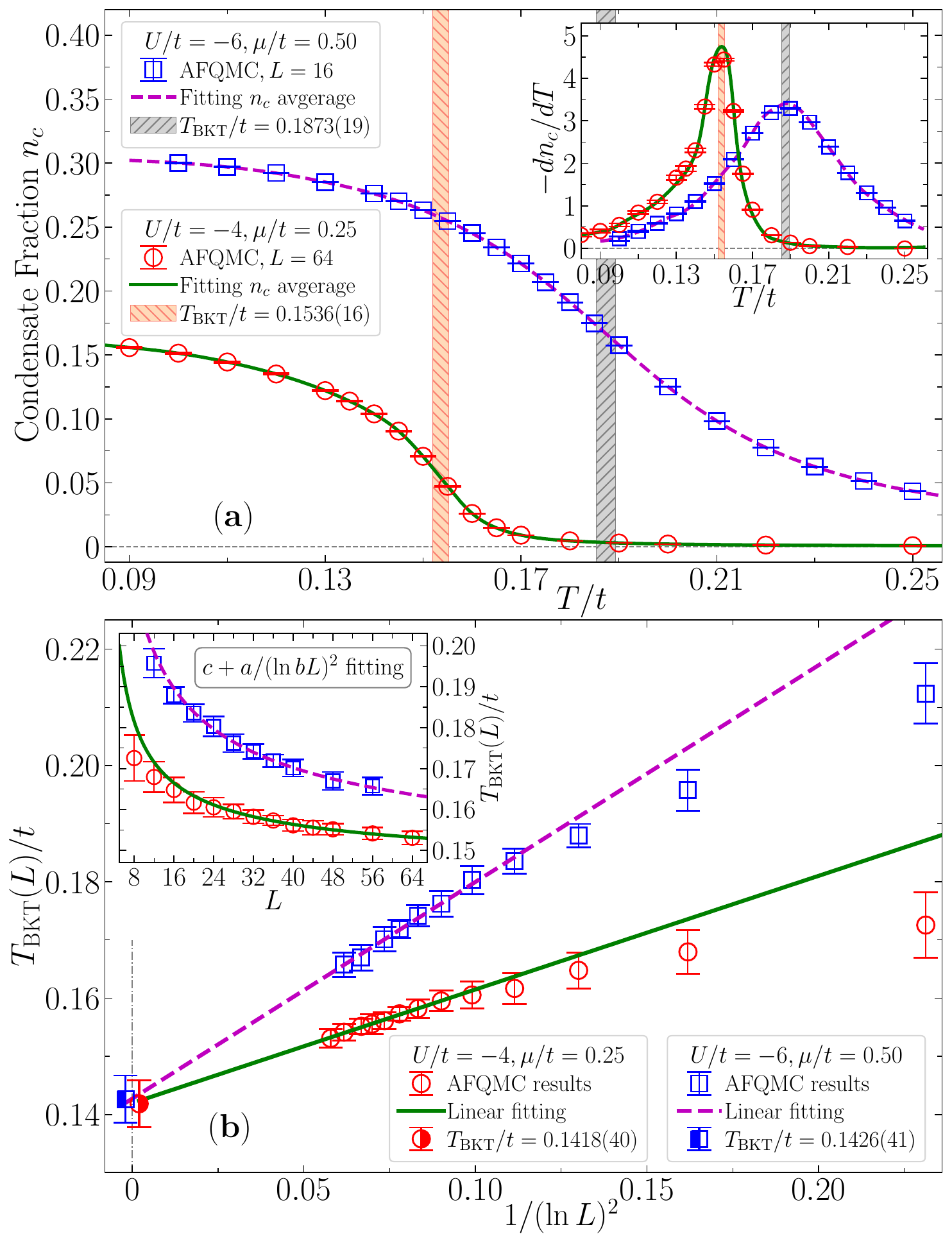}
\caption{Finite-size BKT transition temperature $T_{\rm BKT}(L)$ and its logarithmic correction scaling for the 2D attractive Fermi-Hubbard model. Panel (a) illustrates the determination of $T_{\rm BKT}(L)$ as the inflection point of condensate fraction $n_c(T)$ curve. A cubic-spline fit is first applied to $n_c(T)$, and $T_{\rm BKT}(L)$ is identified as the peak position of $-dn_c/dT$, with error bars estimated via bootstrapping~\cite{Song2024}. The inset plots $-dn_c/dT$, with symbols as results from numerical differentiation~\cite{Chartrand2011,Wietek2021}. Panel (b) shows the scaling of $T_{\rm BKT}(L)$ using $T_{\rm BKT}(L)=T_{\rm BKT}(L=\infty)+a/(\ln L)^2$, and the results of $T_{\rm BKT}(L=\infty)/t$ are also included. The inset plots the fitting of $T_{\rm BKT}(L)$ using a more compact formula of $c+a/(\ln bL)^2$. }
\label{fig:Fig03GetTBKT}
\end{figure}

An alternative way to compute BKT transition temperature is to first obtain the finite-size result $T_{\rm BKT}(L)$ and then extrapolate it to TDL. They satisfy a logarithmic correction relation as $T_{\rm BKT}(L)=T_{\rm BKT}(L=\infty)+a/(\ln bL)^2$~\cite{Ueda2021,Filinov2010,Tomita2002,Nguyen2021,Wenyu2023,Sandvik2010}. This method has been applied in numerical simulations of Ising model~\cite{Wenyu2023}, XY model~\cite{Tomita2002,Nguyen2021}, and dipolar boson system~\cite{Filinov2010}. 

The change from algebraic to exponential scaling behaviors of condensate fraction across the BKT transition naturally gives rise to an inflection point in its temperature dependence, $n_c(T)$. Here we identify this inflection point as $T_{\rm BKT}(L)$. As illustrated in Fig.~\ref{fig:Fig03GetTBKT}(a) for two representative sets of AFQMC data, we first perform fitting for $n_c(T)$ and then determine the inflection point as the peak location of $-dn_c/dT$ using the fitting curve. The uncertainty of $T_{\rm BKT}(L)$ is estimated via bootstrapping~\cite{Song2024}. The consistent results of $-dn_c/dT$ from the fitting and the numerical differentiation using total-variation regularization~\cite{Chartrand2011,Wietek2021} in the inset of Fig.~\ref{fig:Fig03GetTBKT}(a) validates our procedure of computing $T_{\rm BKT}(L)$. We repeat this procedure for $L=8$$\sim$$64$, and perform the scaling for $T_{\rm BKT}(L)$ using the above logarithmic relation. Due to limited precision, we fix the parameter $b=1$ in the fitting, as in previous studies~\cite{Filinov2010,Wenyu2023,Sandvik2010}. The results for [$U/t=-4,\mu/t=0.25$] and [$U/t=-6,\mu/t=0.50$] are plotted in Fig.~\ref{fig:Fig03GetTBKT}(b). TDL transition temperatures of $T_{\rm BKT}(L=\infty)/t=0.1418(40)$ and $0.1426(41)$ are reached for these two cases, respectively, with $n\simeq 0.8003$ and $n\simeq 0.5766$ at the transition~\cite{Suppl}. These numbers are consistent with those obtained from the $\eta(T)$ scaling and the data collapse [see Figs.~\ref{fig:Fig01CondScaling}(a) and \ref{fig:Fig02Collapse}(a) for $U/t=-4$, and Ref.~\onlinecite{Suppl} for $U/t=-6$]. The inset of Fig.~\ref{fig:Fig03GetTBKT}(b) shows the $c+a/(\ln bL)^2$ fitting for $T_{\rm BKT}(L)$, which yields noticeably noisier estimates of $T_{\rm BKT}(L = \infty)$ compared to the fixed $b = 1$ case, but remains consistent within the uncertainty. We have also tried polynomial fits in $1/L$ for $T_{\rm BKT}(L)$, which all produce $T_{\rm BKT}(L = \infty)$ higher than the corresponding bounded range. It thereby supports the validity of the logarithmic correction form instead.

\begin{figure}[t]
\centering
\includegraphics[width=0.960\columnwidth]{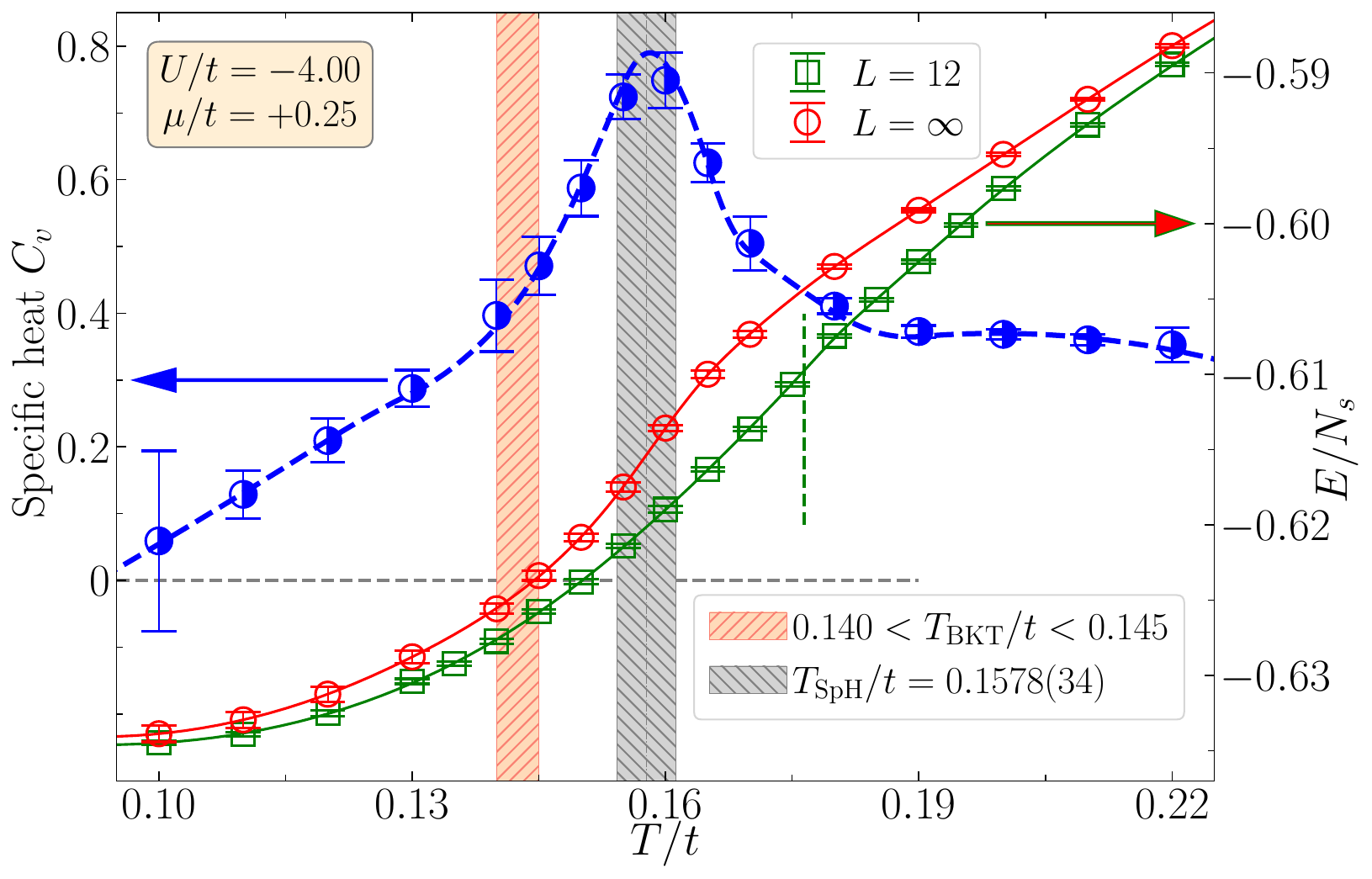}
\caption{Specific heat $C_v$ and total energy $E/N_s$ versus $T/t$ for $U/t=-4,\mu/t=0.25$. Energy results of $L=12$ (green squares) and $L=\infty$ (red circles) with solid lines as polynomial fittings are plotted. TDL results for $C_v$ from the fitting curve of $L=\infty$ energies (blue dash line) and numerical differentiation~\cite{Chartrand2011,Wietek2021} (blue circles) are shown. The peak location of $d(E/N_s)/dT$ for $L=12$ is marked by vertical green dash line as a comparison. The bounded range of $T_{\rm BKT}/t$ and the peak location of specific heat $T_{\rm SpH}/t$ (with the error bar estimated via bootstrapping~\cite{Song2024}) are indicated by the shading bands, leading to the anomaly as $T_{\rm SpH}/T_{\rm BKT}$$\sim$$1.10$.}
\label{fig:Fig04Specific}
\end{figure}

Highly accurate BKT transition temperature can have broad applications in uncovering the underlying physics of 2D correlated systems. One example is the behavior of thermodynamics near the BKT transition. Among these, specific heat is a particularly important and widely used observable for detecting phase transitions~\cite{Mark2012}. However, previous studies showed that the specific heat exhibits a peak at $\sim$$1.6T_{\rm BKT}$ in 2D $^4{\rm He}$~\cite{Ceperley1989} and $\sim$$1.17T_{\rm BKT}$ in 2D XY model~\cite{Tobochnik1979,Ota1992,Ota1995,Gupta1992,Nguyen2021}. Our results also confirm the existence of such anomaly in the 2D attractive Hubbard model. We compute the specific heat using $C_v=(dE/dT)/N_s$ with $E=\langle\hat{H}\rangle$ as the total energy. Through fitting $E/N_s$, the energy derivative is evaluated analytically from the fitting curve. Results of $E/N_s$ ($L=12$, and $L=\infty$ from extrapolations) and $C_v$ (at TDL) are presented in Fig.~\ref{fig:Fig04Specific} for $U/t=-4,\mu/t=0.25$. The specific heat $C_v$ obtained consistently from both energy fitting and numerical differentiation~\cite{Chartrand2011,Wietek2021} exhibits a pronounced peak at $T_{\rm SpH}$$\sim$$1.10T_{\rm BKT}$. Similar behavior is observed in calculations at other fixed values of $\mu/t$, with comparable ratios of $T_{\rm SpH}/T_{\rm BKT}$ and a notable trend of decreasing peak height in $C_v$ with reduced filling~\cite{Suppl}. This specific heat anomaly is attributed to the maximal rate of vortex formation which costs energy, as revealed in 2D XY model~\cite{Tobochnik1979,Ota1992,Ota1995}, and is therefore not directly related to the BKT transition. Nevertheless, it may serve as a precursor signature of approaching BKT transition in future optical lattice experiments for the 2D attracive Hubbard model, 
for which the BKT physics remains inaccessible~\cite{Hartke2023}. 

We have uncovered the scaling behaviors of condensate fraction and developed schemes based on this quantity to estimate BKT transition temperature in the 2D attractive Fermi-Hubbard model. These behaviors and approaches can be readily extended to other 2D correlated fermion systems that exhibit superconductivity or superfluidity, including bilayer models~\cite{Zujev2014,Fontenele2024,Prasad2014,Prasad2022,Prasad2024}, systems with spin-orbit coupling (SOC)~\cite{Shihao2016,Rosenberg2017,Song2024}, and interacting Fermi gases~\cite{Shihao2015,YuanYao2022}. In the presence of SOC, condensate fraction can be similarly computed from a modified pairing matrix $M_{\mathbf{k}\mathbf{k}^{\prime}}$, which incorporates both spin-singlet and spin-triplet pairing~\cite{Song2024,CondnstNote}. Accordingly, the pairing correlators suffer the same issues as shown in Fig.~\ref{fig:Fig01CondScaling}(b) and the accompanying text, while the logarithmic scaling of $T_{\rm BKT}(L)$ derived from condensate fraction proves to be an efficient way to obtain $T_{\rm BKT}$ at TDL~\cite{Song2024}. In a 2D Fermi gas, the dilute limit $n\to0$ renders the pairing correlator and superfluid density to exhibit vanishing signals, even in the finite system simulations. As a result, condensate fraction becomes the most reliable observable for extracting the transition temperature. To obtain the final $T_{\rm BKT}$ (in unit of Fermi temperature $T_F$), a two-step extrapolation procedure for $T_{\rm BKT}(L,N_e)$ sequentially towards $L\to\infty$ and $N_e\to\infty$ is required. As shown in our previous work~\cite{YuanYao2022}, $T_{\rm BKT}(L,N_e)$ exhibits a linear dependence on $1/L^2$ at fixed $N_e$, while the extrapolated value $T_{\rm BKT}(L=\infty,N_e)$ follows a logarithmic scaling with $N_e$. 

Our findings should also be applicable to 2D bosonic systems hosting a superfluid phase, inlcuding the Bose-Hubbard model~\cite{Ohgoe2012}, thin film of $^4$He~\cite{Massimo2006}, and dipolar boson system~\cite{Filinov2010}. In such systems, the condensate fraction can be extracted from the single-particle density matrix $\rho(r)=\langle b_{\mathbf{0}}^+ b_{\mathbf{r}}\rangle$~\cite{Yang1962,Ceperley1995}, which can be easily calculated in various many-body numerical approaches. Thus, our proposed schemes of computing $T_{\rm BKT}$ via condensate fraction provide at least competitive alternatives to the widely used method based on the superfluid density~\cite{Massimo2006,Filinov2010} for many-boson systems.

In summary, we have demonstrated that the {\it condensate fraction} captures the essential physics of BKT transition, exhibiting algebraic and exponential size scalings below and above the the transition, respectively. It also allows to determine the transition temperature with several efficient schemes. Our discoveries can be applied to the 2D superconducting and superfluid transitions in a broad class of correlated fermion and boson systems. In the 2D attractive Fermi-Hubbard model, our results show that the condensate fraction scaling exhibits remarkably smaller finite-size effect than the commonly used on-site pairing correlator, and thus can substantially change the prospect of computing BKT transition temperature. We further extract the finite-size transition temperature from the condensate fraction in the model, and confirm its expected logarithmic scaling behavior. Additionally, we compute the specific heat and observe its anomaly in this fermionic system, which may offer valuable guidance for optical lattice experiments aiming to probe BKT transition in attractive fermion systems.

{\it{Acknowledgements}}. The author thanks Youjin Deng, Massimo Boninsegni, and Shiwei Zhang for insightful discussions on the finite-size scaling and applications to bosonic systems. This work was supported by the National Natural Science Foundation of China (under Grants No. 12247103 and No. 12204377), the Quantum Science and Technology-National Science and Technology Major Project (Grant No. 2021ZD0301900), and the Youth Innovation Team of Shaanxi Universities.

\bibliography{Hubbard_BKTMain}


\onecolumngrid
\newpage

\begin{center}
\textbf{\large Supplementary Material for \\
``Condensate Fraction Scaling and Berezinskii-Kosterlitz-Thouless Transition of Superconductivity and Superfluidity''}
\end{center}

\section{I. Overview of this supplementary material}
\label{sec:Overview}

In this Supplementary Material, we present all the additional information for the main text, including the algorithmic progresses and all the other numerical results from our AFQMC simulations. 

In 
Sec. II, we present a summary of the lattice model, a detailed description for the important improvements of AFQMC method applied in our simulations, physical observables, and the AFQMC simulation setup in this work. 

In 
Sec. III, we present additional AFQMC results for fixed $\mu/t$ calculations, including $U/t=-4$ with $\mu=0.25,0.60,1.25$ and $U/t=-6$ with $\mu=0.5$. For $U/t=-4,\mu/t=0.6$ case, we have also computed the superfluid density and determined the BKT transition temperature.

In 
Sec. IV, we present additional AFQMC results for fixed fermion filling calculations, including $U/t=-4$ and $U/t=-10$ with fermion filling $\langle\hat{n}\rangle=0.50$.

All these supplementary results mainly concentrate on the temperature dependence and size scaling of condensation fraction and on-site pairing correlator, and total energy and specific heat. To further establish the scheme of characterizing the BKT transition by condensate fraction, we have also calculated and presented results of superfluid density for representative parameter set in 
Sec. III.

\section{II. Methodology}
\label{sec:modelmethod}

We first introduce the 2D attractive Hubbard model in Sec. II A, and then in Sec. II B present the finite-temperature AFQMC method with important improvements that we have applid in this work to numerically solve the model. Then we define the physical observables we have calculated related to the BKT physics in Sec. II C. Next, we analyze the finite-size effect in the on-site pairing correlator $\langle\Delta^2\rangle$ in Sec. II D. Finally, we decribe the AFQMC simulation setup in Sec. II E.

\subsection{A. The lattice model}
\label{sec:Model}

We study the attracive Hubbard model on 2D square lattice with the Hamiltonian as
\begin{equation}\begin{aligned}
\label{eq:ModelHmtSup}
\hat{H} = 
-t\sum_{\langle\mathbf{ij}\rangle\sigma}(c_{\mathbf{i}\sigma}^+c_{\mathbf{j}\sigma}+c_{\mathbf{j}\sigma}^+c_{\mathbf{i}\sigma}) + \mu\sum_{\mathbf{i}} (\hat{n}_{\mathbf{i}\uparrow} + \hat{n}_{\mathbf{i}\downarrow} ) 
+ U\sum_{\mathbf{i}}\Big( \hat{n}_{\mathbf{i}\uparrow}\hat{n}_{\mathbf{i}\downarrow} - \frac{\hat{n}_{\mathbf{i}\uparrow} + \hat{n}_{\mathbf{i}\downarrow}}{2} \Big),
\end{aligned}\end{equation}
with $\hat{n}_{\mathbf{i}\sigma}=c_{\mathbf{i}\sigma}^+ c_{\mathbf{i}\sigma}$ as the density operator for spin-$\sigma$ channel (with $\sigma=\uparrow,\downarrow$). We only consider the nearest-neighbor hopping $t$ with the energy dispersion as $\varepsilon_{\mathbf{k}}=-2t(\cos k_x+\cos k_y)$, where $k_x,k_y$ are the momentum defined in units of $2\pi/L$ with the linear system size $L$. The fermion filling reads $n=N/N_s$ with $N_s=L^2$ and $N$ as the number of lattice sites and fermions in the system. Other tunning model paremeters are the on-site Coulomb interaction $U$ and chemical potential $\mu$. Thoughout this work, we set $t$ as the energy unit, and focus on attractive interaction as $U<0$ and hole doping regime with $\mu>0$. 

It is well-known that the lattice model in Eq.~(\ref{eq:ModelHmtSup}) with $U<0$ has a superconducting ground state with spin-singlet $s$-wave pairing order for arbitrary filling~\cite{Scalettar1989,Singer1996}. At half filling ($\mu=0$), besides the spin symmetry, the model also has charge SU(2) symmetry~\cite{Yang1989,Mark2020,Hille2020} which has generators $\hat{\bf{N}}=(\hat{\eta}^+, \hat{\eta}, \hat{\eta}^z)$ with $\hat{\eta}^+=\sum_{\bf{i}}(-1)^{\bf{i}}c_{\bf{i}\uparrow}^+c_{\bf{i}\downarrow}^+$ and $\hat{\eta}^z=\frac{1}{2}\sum_{\bf{i}}(\hat{n}_{\bf{i}\uparrow}+\hat{n}_{\bf{i}\downarrow}-1)$. The first two components stands for the pairing order with the third one corresponding to the charge density wave (CDW) order, and these orders are degenerate as a result of the symmetry. With doping ($\mu\ne0$), the chemical potential term breaks the charge SU(2) symmetry to U(1) (charge conservation) by lifting the energy of CDW ordered state while the superconductivity remains unaffected as the ground state~\cite{Mark2020}. As a result, the system acquires the BKT transition of the superconducting order at finite temperature~\cite{Paiva2004,Fontenele2022}, which is the main content of this work. 

\subsection{B. The finite-temperature AFQMC method}
\label{sec:AFQMC}

The key idea of AFQMC methods~\cite{Blankenbecler1981,Hirsch1983,White1989,YuanYao2019,Yuanyao2019b} is to apply the Hubbard-Stratonovich (HS) transformation~\cite{Hirsch1983} to decouple the two-body interaction into free fermion operators coupled with classical fields, which then allows for calculations of fermionic observables via the imporatance sampling of auxiliary-field configurations~\cite{White1989}. It has been widely applied to variously correlated fermion systems in condensed matter physics. Specially, the attractive Hubbard model in Eq.~(\ref{eq:ModelHmtSup}) is free of sign problem for arbitrary filling in AFQMC simulations. In the following, we present important algorithmic details applied in our numerical simulations. 

The finite-temperature AFQMC algorithm starts from the imaginary-time discretization as $\beta=M\Delta\tau$, and expresses the partition function as $Z=\text{Tr}(e^{-\beta\hat{H}})=\text{Tr}[(e^{-\Delta\tau\hat{H}})^M]$. Then the symmetric Trotter-Suzuki decomposition as $e^{-\Delta\tau\hat{H}}=e^{-\Delta\tau\hat{H}_0/2}e^{-\Delta\tau\hat{H}_I}e^{-\Delta\tau\hat{H}_0/2}+\mathcal{O}[(\Delta\tau)^3]$ is applied to seperate the interaction term $\hat{H}_I$ from free fermion part $\hat{H}_0$. The Trotter error $\mathcal{O}[(\Delta\tau)^3]$ is typically eliminated by the extrapolation to $\Delta\tau\to0$ limit, and thus it is neglected in the following. As for the HS transformation, the practical choices for the attractive Hubbard interaction ($U<0$) are the discrete ones~\cite{Hirsch1983} into charge density channel as
\begin{equation}\begin{aligned}
\label{eq:HSdensity}
\exp\Big[-\Delta\tau U\Big(\hat{n}_{\mathbf{i}\uparrow}\hat{n}_{\mathbf{i}\downarrow} - \frac{\hat{n}_{\mathbf{i}\uparrow} + \hat{n}_{\mathbf{i}\downarrow}}{2}\Big)\Big]
= \frac{e^{+\Delta\tau U/2}}{2} \sum_{x_{\mathbf{i}}=\pm1}e^{\gamma x_{\mathbf{i}}(\hat{n}_{\mathbf{i}\uparrow} + \hat{n}_{\mathbf{i}\downarrow}-1)},
\end{aligned}\end{equation}
with the real coupling constant $\gamma=\cosh^{-1}(e^{-\Delta\tau U/2})$, and into spin-$\hat{s}_z$ channel as
\begin{equation}\begin{aligned}
\label{eq:HSSz}
\exp\Big[-\Delta\tau U\Big(\hat{n}_{\mathbf{i}\uparrow}\hat{n}_{\mathbf{i}\downarrow} - \frac{\hat{n}_{\mathbf{i}\uparrow} + \hat{n}_{\mathbf{i}\downarrow}}{2}\Big)\Big]
= \frac{1}{2} \sum_{x_{\mathbf{i}}=\pm1} e^{\gamma x_{\mathbf{i}}(\hat{n}_{\mathbf{i}\uparrow} - \hat{n}_{\mathbf{i}\downarrow})},
\end{aligned}\end{equation}
with $\gamma={\rm i}\cos^{-1}(e^{+\Delta\tau U/2})$ as an imaginary number. Although the formula in Eq.~(\ref{eq:HSdensity}) only involves real quantities in the whole AFQMC simulations, it explicitly breaks the charge SU(2) symmetry of the Hubbard interaction term as mentioned in Sec. II A, which as a result usually causes strong fluctuations and large statistical errors for pairing and charge density related quantities regardless of half-filling or doping. Instead, the HS transformation in Eq.~(\ref{eq:HSSz}) preserves the charge SU(2) symmetry and can present precise results for pairing observables in the Monte carlo calculations. Thus, we always adopt the HS transformation into spin-$\hat{s}_z$ channel as Eq.~(\ref{eq:HSSz}) in this work. After all the above procedures, we can reach the form $e^{-\Delta\tau\hat{H}} = \sum_{\mathbf{x}_{\ell}}p(\mathbf{x}_{\ell})\hat{B}_{\ell}$ [with $p(\mathbf{x}_{\ell})=1/2^{N_s}$ within Eq.~(\ref{eq:HSSz})] for the $\ell$-th time slice, where the single-particle propagator $\hat{B}_{\ell}$ reads
\begin{equation}\begin{aligned}
\label{eq:BlOperator}
\hat{B}_{\ell}
= e^{-\Delta\tau\hat{H}_0/2} \Big[ \prod_{\mathbf{i}}e^{\gamma x_{\ell,\mathbf{i}}(\hat{n}_{\mathbf{i}\uparrow} - \hat{n}_{\mathbf{i}\downarrow})} \Big] e^{-\Delta\tau\hat{H}_0/2}
= e^{-\Delta\tau\hat{H}_0/2} e^{\hat{V}(\mathbf{x}_{\ell})}
e^{-\Delta\tau\hat{H}_0/2},
\end{aligned}\end{equation}
with $\mathbf{x}_{\ell}=\{x_{\ell,1},x_{\ell,2},\cdots,x_{\ell,\mathbf{i}},\cdots,x_{\ell,N_s}\}$ and $\hat{V}(\mathbf{x}_{\ell})=\sum_{\mathbf{i}}\gamma x_{\ell,\mathbf{i}}(\hat{n}_{\mathbf{i}\uparrow} - \hat{n}_{\mathbf{i}\downarrow})$. Then the partition function can be evaluated as $Z=\sum_{\mathbf{X}}W(\mathbf{X})$ with the configuration weight
\begin{equation}\begin{aligned}
\label{eq:weight}
W(\mathbf{X}) 
= P(\mathbf{X})\cdot\text{Tr}(\hat{B}_M\hat{B}_{M-1}\cdots\hat{B}_2\hat{B}_1)
= P(\mathbf{X})\prod_{\sigma=\uparrow,\downarrow} \det(\mathbf{1}_{N_s}+\mathbf{B}_M^{\sigma}\cdots\mathbf{B}_2^{\sigma}\mathbf{B}_1^{\sigma}),
\end{aligned}\end{equation}
where $\mathbf{X}=\{\mathbf{x}_1,\mathbf{x}_2,\cdots,\mathbf{x}_{\ell},\cdots,\mathbf{x}_{M}\}$ is a complete configuration (or path) in auxiliary-field space, and $P(\mathbf{X})=1/2^{N_sM}$ within Eq.~(\ref{eq:HSSz}). The second equality in Eq.~(\ref{eq:weight}) has applied the property that the $\hat{B}_{\ell}$ operator is spin-decoupled, and the corresponding propagation matrix has the form $\mathbf{B}_{\ell}^{\sigma}=e^{-\Delta\tau\mathbf{H}_0^{\sigma}/2}e^{\mathbf{H}_I^{\sigma}(\mathbf{x}_{\ell})}e^{-\Delta\tau\mathbf{H}_0^{\sigma}/2}$, with $\mathbf{H}_0^{\sigma}$ and $\mathbf{H}_I^{\sigma}(\mathbf{x}_{\ell})=(-1)^{\sigma}\gamma\cdot{\rm Diag}(x_{\ell,1},x_{\ell,2},\cdots,x_{\ell,N_s})$ as the $N_s\times N_s$ hopping matrix for $\hat{H}_0$ and the representation matrix for $\hat{V}(\mathbf{x}_{\ell})$ in Eq.~(\ref{eq:BlOperator}), respectively. The relation $\mathbf{B}_{\ell}^{\downarrow}=(\mathbf{B}_{\ell}^{\uparrow})^{\star}$ apparently holds, and the spin-up and -down determinants in Eq.~(\ref{eq:weight}) are complex conjugate, resulting in $W(\mathbf{X})\ge0$ as free of sign problem. Following the above formalism, the imporatance sampling of auxiliary-field configurations is then performed by Markov Chain process with Metropolis algorithm. The physical observables for fermions are first measured from single-particle Green's function for chosen configurations, and then collected to account for statistics. 

\begin{figure}[ht!]
\centering
\includegraphics[width=0.55\columnwidth]{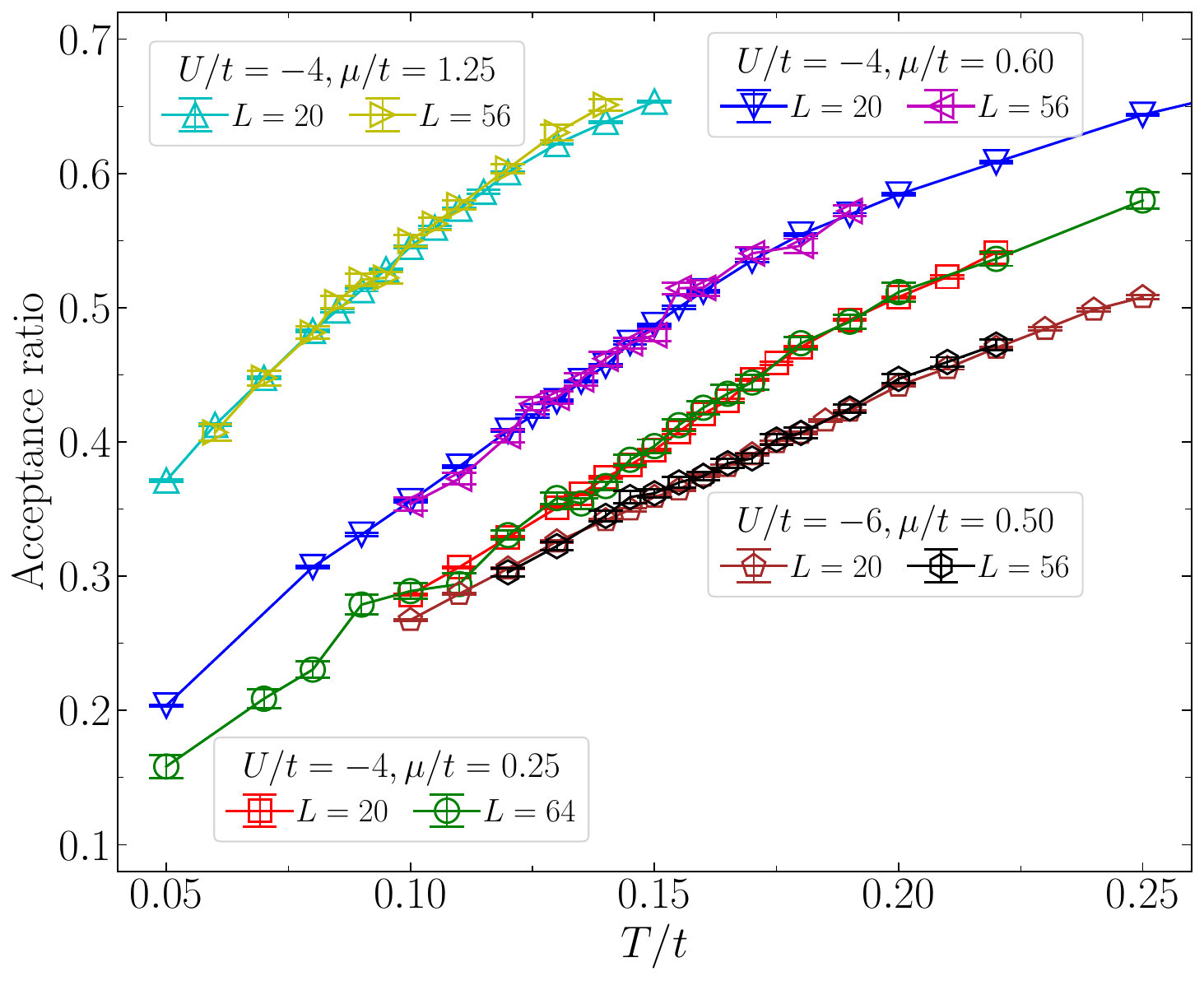}
\caption{The acceptance ratio of the $\tau$-line global update in AFQMC simulations for four different sets of parameters. The ratio typically increases towards the low filling regime (with larger $\mu/t$), and it decreases with stronger interaction. }
\label{fig:Fig01GlobalUpdate}
\end{figure}

Several key techniques and improvements are applied in our AFQMC simulations. {\it First}, we apply the Fast Fourier Transform (FFT) in the propagation (sequential product) of $\mathbf{B}_{\ell}^{\sigma}$ matrices. Within the unitary transformation matrix $\mathbf{U}$ between $\mathbf{r}$- and $\mathbf{k}$-space, we arrive at $e^{-\Delta\tau\mathbf{H}_0^{\sigma}/2}=\mathbf{U}\mathbf{U}^+e^{-\Delta\tau\mathbf{H}_0^{\sigma}/2}\mathbf{U}\mathbf{U}^+=\mathbf{U}e^{-\Delta\tau\boldsymbol{\Lambda}/2}\mathbf{U}^+$ with $\boldsymbol{\Lambda}={\rm Diag}(\varepsilon_{\mathbf{k}_1},\varepsilon_{\mathbf{k}_2},\cdots,\varepsilon_{\mathbf{k}_{N_s}})$, where $\varepsilon_{\mathbf{k}}$ is the band dispersion defined in Sec. II A. Then we have $\mathbf{B}_{\ell}^{\sigma}=\mathbf{U}e^{-\Delta\tau\boldsymbol{\Lambda}/2}\mathbf{U}^+e^{\mathbf{H}_I^{\sigma}(\mathbf{x}_{\ell})}\mathbf{U}e^{-\Delta\tau\boldsymbol{\Lambda}/2}\mathbf{U}^+$, which explicitly modifies the propagation of $\mathbf{B}_{\ell}^{\sigma}$ matrix by multiplying diagonal matrices in both $\mathbf{r}$- and $\mathbf{k}$-space plus several FFT operations. This technique implements the diagonal properties of both $\mathbf{H}_0^{\sigma}$ in $\mathbf{k}$-space (under periodic boundary conditions) and $\mathbf{H}_I^{\sigma}(\mathbf{x}_{\ell})$ in $\mathbf{r}$-space (with local interactions). In practical simulations, the FFT operations are realized with standard library (such as FFTW). The computational effort of the matrix propagations can then be reduced to $\mathcal{O}(\beta N_s^2\log N_s)$ from the conventional $\mathcal{O}(\beta N_s^3)$ scaling. {\it Second}, we implement the delayed update~\cite{McDaniel2017} and single $\tau$-line global update~\cite{Scalettar1991}. The former is an accelerated way of doing local update, while the latter is a carefully designed cluster update to alleviate the possible ergodicity problem in AFQMC simulations. Delayed update applies rank-$K$ Sherman-Morrison updating scheme (with $K$ as a predetermined number) instead of the rank-1 formula used in local update, and the acceleration originates from matrix-matrix operation over $K$ times of individual matrix-vector operations. This procedure does not alter the Monte Carlo sampling or its statistical efficiency, but instead it reduces the prefactor of the computational scaling [as $\mathcal{O}(\beta N_s^3)$] of local update. For every $\tau$-line global update, we simply choose a lattice site $\mathbf{i}$ and flip all the auxiliary fields $\{x_{\ell,\mathbf{i}}, \ell=1,2,\cdots,M\}$ along the imaginary time direction. Figure~\ref{fig:Fig01GlobalUpdate} plots the overall acceptance ratio of this global update for different simulation parameters. We can see that it can still reach $0.16$ for $64\times 64$ system with $U/t=-4$ and $T/t=0.05$, and it further grows for higher temperature and stronger interaction. Thus, this global update surely improves the efficiency of the imporatance sampling. Pratically, one Monte Carlo sweep in AFQMC simulations contains $\beta\to0$ sweep of local updates, $\tau$-line global update, and $0\to\beta$ sweep of local updates. ${\it Third}$, we use column-pivoted QR algorithm for stabilizing the matrix products, and successively a highly stable formalism to obtain the static single-particle Green's function matrix $\mathbf{G}^{\sigma}(\tau,\tau)=\{G_{\mathbf{ij}}^{\sigma}=\langle c_{\mathbf{i}\sigma}c_{\mathbf{j}\sigma}^+\rangle_{\tau}\}$ (with details presented in Ref.~\onlinecite{Yuanyao2019b}). Moreover, we compute $\mathbf{G}^{\sigma}(\tau,\tau)$ (for a specific configuration) at $\tau=\ell\Delta\tau$ from scratch after the numerical stabilization procedure as
\begin{equation}\begin{aligned}
\label{eq:GMatrix}
\mathbf{G}^{\sigma}(\tau,\tau)=
\big(\mathbf{1}_{N_s}+e^{-\beta\mu}\mathbf{B}_{\ell}^{\sigma}\cdots\mathbf{B}_{1}^{\sigma}\mathbf{B}_{M}^{\sigma}\cdots\mathbf{B}_{\ell+1}^{\sigma}\big)^{-1}.
\end{aligned}\end{equation}
The chemical potential term $e^{-\beta\mu}$ is seperated from the hopping terms in $\mathbf{H}_0^{\sigma}$ matrix to relieve the ill condition of the matrices, and make the numerical simulations more stable. This improvement is quite important for low filling systems with large $\mu$, and it allows for AFQMC simulations at lower temperatures. {\it Fourth}, for dilute systems, we can apply our recent progress in AFQMC~\cite{YuanYao2019} as the speedup algorithm with overall computational scaling $\mathcal{O}(\beta N_s N^2)$. 

All the above techniques largely extends the capabilty of finite-temperature AFQMC method, and allow us to obtain results with satisfying precision of $64\times 64$ system for the 2D Hubbard model in Eq.~(\ref{eq:ModelHmt}) using moderate computational resource. For example, such system with $U/t=-4,\mu/t=0.25$ and $T/t=0.05$ costs $\sim$$2\times 10^4$ CPU hours, which produces the results with relative errors $\sim$$0.5\%$ for the condensate fraction and on-site pairing correlator. Besides, all these improvements can be easily implemented in AFQMC simulations of other more complicated systems, such as the spin-orbit coupled Hubbard model~\cite{Song2024}. 

\subsection{C. Physical observables}
\label{sec:PhysObs}

Condensate fraction is the key quantity we use to characterize the pairing properties as well as the BKT transition in this work. It can be computed from the pairing matrix in momentum space~\cite{Shihao2015,YuanYao2019,YuanYao2022} as
\begin{equation}\begin{aligned}
\label{eq:PairMat}
M_{\mathbf{k}\mathbf{k}^{\prime}}
=\langle\hat{\Delta}_{\mathbf{k}}^+\hat{\Delta}_{\mathbf{k}^{\prime}}\rangle-\delta_{\mathbf{k}\mathbf{k}^{\prime}}\langle c_{\mathbf{k}\uparrow}^+c_{\mathbf{k}\uparrow}\rangle\langle c_{-\mathbf{k}\downarrow}^+c_{-\mathbf{k \downarrow}}\rangle,
\end{aligned}\end{equation}
with $\hat{\Delta}_{\mathbf{k}}^+=c_{\mathbf{k}\uparrow}^+c_{-\mathbf{k}\downarrow}^+$ denoting the $\mathbf{k}$-space spin-singlet pairing operator with {\it zero center-of-mass momentum}. Thus, the above pairing matrix can be taken as the dominant component of the two-body density matrix $\langle c_{\mathbf{i}\uparrow}^+c_{\mathbf{j}\downarrow}^+c_{\mathbf{i}+\mathbf{r}\uparrow}c_{\mathbf{j}+\mathbf{r}^{\prime}\downarrow}\rangle$ related to the off-diagonal long-range order~\cite{Yang1962,Boronat2005} for attractive Hubbard model. Its leading eigenvalue $\lambda_{\rm max}$ is associated with the condensate fraction as $n_c=\lambda_{\rm max}/(N/2)$, meaning the fraction of Bose-condensed fermion pairs. The corresponding eigenstate of $\lambda_{\rm max}$ is the momentum-space pair wave function $\phi_{\uparrow\downarrow}(\bf{k})$ whose square identifies the probability of fermions with momentum $\bf{k}$ participating the pairing. Its Fourier transform $\psi_{\uparrow\downarrow}(\bf{r})$ reflects the size distribution of Cooper pairs. In AFQMC simulations, we measure $M_{\mathbf{k}\mathbf{k}^{\prime}}$ for a specific configuration applying Wick decomposition directly from the $\mathbf{k}$-space single-particle Green's function matrix $\mathbf{G}^{\sigma}=\{G_{\mathbf{k},\mathbf{k}^{\prime}}^{\sigma}=\langle c_{\mathbf{k}\sigma}c_{\mathbf{k}^{\prime}\sigma}^+\rangle_{\tau}\}$, which is calculated by FFT of the $\mathbf{r}$-space $\mathbf{G}^{\sigma}(\tau,\tau)$ matrix in Eq.~(\ref{eq:GMatrix}).

The other quantities for the pairing pheomena is the pairing correlation function and structure factor for on-site spin-singlet Cooper pairs, which is widely studied in previous work~\cite{Paiva2004,Fontenele2022}. Here we define the pairing operator as $\hat{\Delta}_{\mathbf{i}}^+=c_{\mathbf{i}\uparrow}^+c_{\mathbf{i}\downarrow}^+$, and then the correlation function and the corresponding structure factor can be expressed as $P(\mathbf{r})=\langle\hat{\Delta}_{\mathbf{i}}^+\hat{\Delta}_{\mathbf{i}+\mathbf{r}}+\hat{\Delta}_{\mathbf{i}}\hat{\Delta}_{\mathbf{i}+\mathbf{r}}^+\rangle/4$ and $S(\mathbf{q})=\sum_{\mathbf{r}}P(\mathbf{r})e^{i\mathbf{q}\cdot\mathbf{r}}$. From $S(\mathbf{q})$, we can identify the on-site pairing correlator as $\langle\Delta^2\rangle=S(\mathbf{q}=\boldsymbol{\Gamma})/L^2$, and $\sqrt{\langle\Delta^2\rangle}$ at $T=0$ is the magnitude of the long-range pairing order for the ground state. The calculation of $\langle\Delta^2\rangle$ in AFQMC is straightforward from the $\mathbf{r}$-space $\mathbf{G}^{\sigma}(\tau,\tau)$ matrix. 

The pairing matrix $M_{\mathbf{k}\mathbf{k}^{\prime}}$ in Eq.~(\ref{eq:PairMat}) formally resembles the pairing structure factor $S(\mathbf{q})$. Physically, Cooper pairs with different sizes including local and nonlocal ones should coexist in attractive Hubbard model at low temperature regime~\cite{Hartke2023}. However, the $\hat{\Delta}_{\mathbf{i}}^+$ operator for $S(\mathbf{q})$ only accounts for the local Cooper pairs, while the $\hat{\Delta}_{\mathbf{k}}^+$ for $M_{\mathbf{k}\mathbf{k}^{\prime}}$ directly probe the $\mathbf{k}$-space pairing and thus contains the information of Cooper pairs with various sizes. As a result, $M_{\mathbf{k}\mathbf{k}^{\prime}}$ togather with the condensate fraction $n_c$ and pair wave function $\phi_{\uparrow\downarrow}(\bf{k})$ should be a more complete way to characterize the pairing properties for the system of attractive fermions.

Another important observable related to superconductivity and superfluidity is the superfluid density $\rho_s$ (also called ``helicity modulus''). It was found~\cite{Scalapino1992,*Scalapino1993} that this quantity can be computed with the dynamic current-current correlation function. The current operator is defined as $\hat{j}_{\mathbf{i}}=it\sum_{\sigma}(c_{\mathbf{i},\sigma}^+c_{\mathbf{i}+\mathbf{e}_x,\sigma} - c_{\mathbf{i}+\mathbf{e}_x,\sigma}^+c_{\mathbf{i},\sigma})$ with $\mathbf{e}_x$ representing the unit lattice space in $x$-direction. Then we can obtain its dynamic correlation as $\Lambda_{\mathbf{ij}}^{xx}(\tau) = \langle\hat{j}_{\mathbf{i}}(\tau)\hat{j}_{\mathbf{j}}(0)\rangle$ with $\hat{j}_{\mathbf{i}}(\tau)=e^{\tau\hat{H}}\hat{j}_{\mathbf{i}}e^{-\tau\hat{H}}$. Then we can reach the imaginary-frequency correlator in momentum space via Fourier transformations as 
\begin{equation}\begin{aligned}
\label{eq:CurrentCrFt}
\Lambda_{xx}(\mathbf{q},i\omega_m) 
= \int_{0}^{\beta} \Lambda_{xx}(\mathbf{q},\tau)e^{i\omega_m\tau} d\tau 
= \frac{1}{N_s} \int_{0}^{\beta} e^{i\omega_m\tau} \sum_{\mathbf{ij}} \Lambda_{\mathbf{ij}}^{xx}(\tau) e^{i\mathbf{q}\cdot(\mathbf{R}_{\mathbf{i}}-\mathbf{R}_{\mathbf{j}})} d\tau,
\end{aligned}\end{equation}
with $\omega_m=2\pi m/\beta$. Then the superfluid density can be expressed as
\begin{equation}\begin{aligned}
\rho_s = \frac{1}{4}\Big[\Lambda_{xx}(q_x\to0,q_y=0,i\omega_m=0) - \Lambda_{xx}(q_x=0,q_y\to0,i\omega_m=0) \Big].
\end{aligned}\end{equation}
The first term in Eq.~(\ref{eq:CurrentCrFt}) actually equals the opposite of kinetic energy associated with the x-oriented links as $\Lambda_{xx}(q_x\to0,q_y=0,i\omega_m=0)=\langle-\hat{K}_x\rangle$. In practical AFQMC calulations, we first measure the real-space dynamic current-current correlation function $\Lambda_{\mathbf{ij}}^{xx}(\tau)$ for each configuration using standard Wick decomposition, and then transform it into the momentum space as $\Lambda_{xx}(\mathbf{q},\tau)$, and finally obtain the imaginary-frequency result $\Lambda_{xx}(\mathbf{q},i\omega_m)$ via another Fourier transformation. For finite system with linear system size $L$, we replace the $q_x\to0$ and $q_y\to0$ in Eq.~(\ref{eq:CurrentCrFt}) by $q_x=2\pi/L$ and $q_y=2\pi/L$ in the calculations. Typically, the results dynamic current-current correlation and thus superfluid density are rather noisy due to the much heavier computational cost of calculating dynamic properties in AFQMC method. In this work, we have reached $L=32$ (with $N_s=1024$) results for the superfluid density.

\subsection{D. Analysis for the finite-size effect of pairing correlator}
\label{sec:PairCrFfs}

\begin{figure}[ht!]
\centering
\includegraphics[width=0.55\columnwidth]{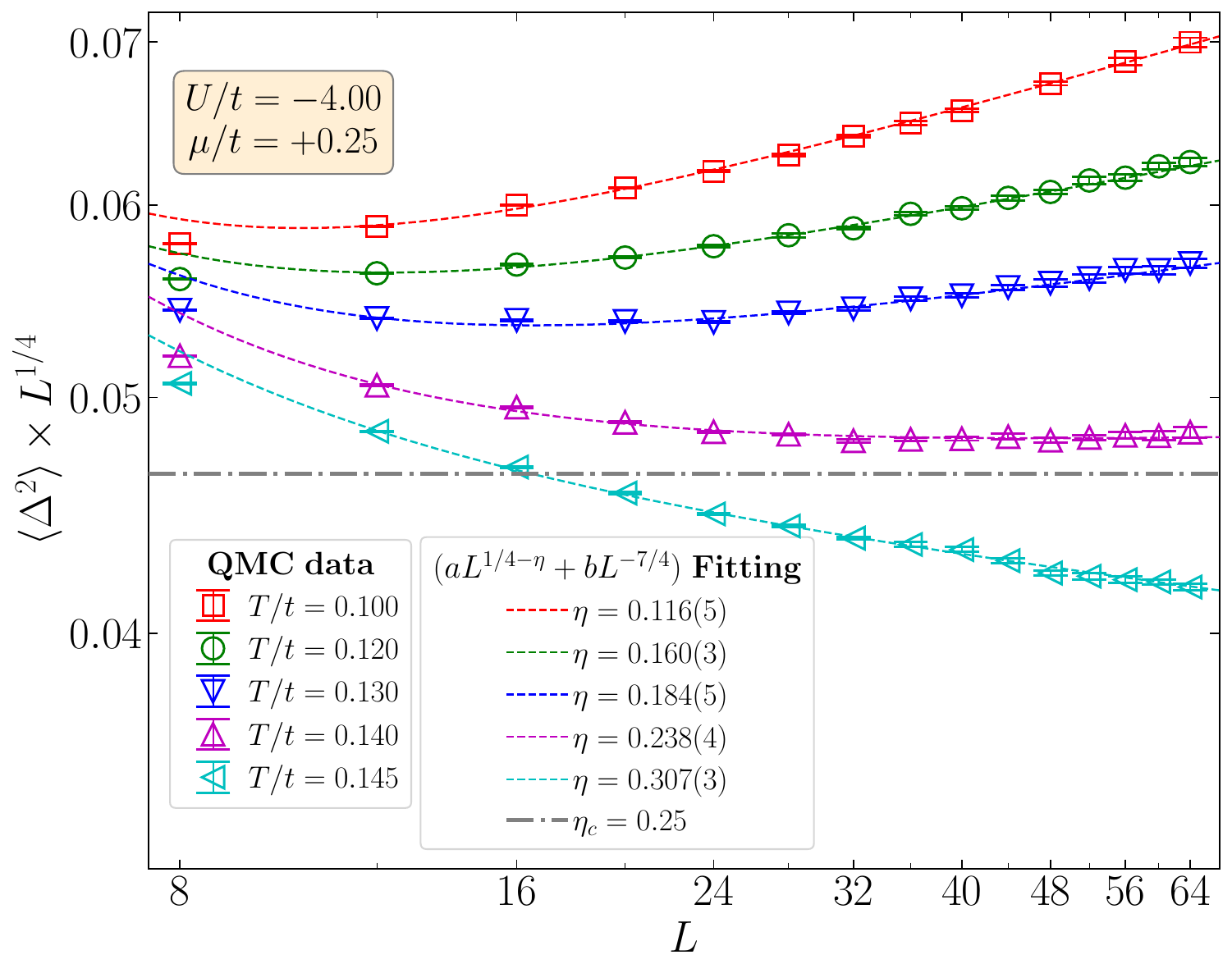}
\caption{
Fitting the numerical results of the rescaled pairing correlator $\langle\Delta^2\rangle\times L^{1/4}$ at $T/t=0.100\sim 0.145$ for $U/t=-4,\mu/t=0.25$, using $(aL^{1/4-\eta}+bL^{-7/4})$ to include the sub-leading term revealed by Eq.~(\ref{eq:DeltaSquareFinl}). This fit works well for $L\ge 12$, and yields consistent results of the exponent $\eta$ with the asymptotic scaling (i.e., $\langle\Delta^2\rangle\propto L^{-\eta}$) at large system size ($L\ge 32$), as shown in Fig. 1 in the main text.
}
\label{fig:FigOrderSquare}
\end{figure}

In superfluid phase, the on-site pairing correlation function $P(\mathbf{r})$, as defined in Sec. II C for 2D systems, should decay algebraically in large $r$ limit, i.e., $P(r)=g(r)=A\times r^{-\eta}$ with $\eta<\eta_c$ at $T<T_{\rm BKT}$ and the critical exponent $\eta_c=1/4$ at the BKT transition point. Note that, within translational symmetry and periodic boundary conditions, $P(\mathbf{r})$ only depends on the distance $r$. For this case, we can decompose $P(r)$ into two seperate regions as $r\ge r_0$ and $r<r_0$, which satisfies
\begin{equation}\begin{aligned}
\label{eq:nrDecomp}
P(r) = \left\{\begin{array}{ll}
	f(r)\ \ {\rm for} \ \ r<r_0  \\
	g(r)\ \ {\rm for} \ \ r\ge r_0
\end{array},
\right.
\end{aligned}\end{equation}
where $f(r)$ is a parameter dependent function. At thermodynamic limit (TDL), the pairing correlator can be expressed as
\begin{equation}\begin{aligned}
\label{eq:DeltaSquare}
\langle\Delta^2\rangle 
&= \frac{1}{L^2}\sum_{\mathbf{r}}P(\mathbf{r})
= \frac{1}{V}\int P(\mathbf{r})d\mathbf{r}
= \frac{1}{L^2}\int_0^{2\pi}d\theta\int_0^{b L}rdr \cdot P(r) \\
&= \frac{2\pi}{L^2}\int_0^{b L}r P(r)dr
 = \frac{2\pi}{L^2}\Big[\int_0^{r_0}rf(r)dr + \int_{r_0}^{b L}rg(r)dr\Big].
\end{aligned}\end{equation}
We have considered the fact that, for a finite system with $V=L^2$, the largest distance should be proportional to $L$, and assign the upper limit of $r$ integral as $b L$ with $b\sim 1$. In the last equality of Eq.~(\ref{eq:DeltaSquare}), the first integral with the integrand $rf(r)$ should produce some contant $C_0$, which should finally be independet of $L$ (with increasing $L$). For the second integral, we can substitute $g(r)=A\times r^{-\eta}$ and evaluate it as
\begin{equation}\begin{aligned}
\int_{r_0}^{b L}rg(r)dr = a\int_{r_0}^{b L}r^{1-\eta}dr
=\frac{(bL)^{2-\eta} - (r_0)^{2-\eta}}{2-\eta}.
\end{aligned}\end{equation}
Then $\langle\Delta^2\rangle$ in Eq.~(\ref{eq:DeltaSquare}) can be simplied as
\begin{equation}\begin{aligned}
\label{eq:DeltaSquareFinl}
\langle\Delta^2\rangle
={\color{red} \frac{2\pi b^{2-\eta}}{2-\eta}L^{-\eta}} + {\color{blue} 2\pi\Big[C_0 - \frac{(r_0)^{2-\eta}}{2-\eta}\Big] L^{-2}}.
\end{aligned}\end{equation}
It is clear that, the first term (in red) is the leading contribution of $\langle\Delta^2\rangle$ ($\propto L^{-\eta}$), while the second term (in blue) is the subleading correction ($\propto L^{-2}$). As a result, Eq.~(\ref{eq:DeltaSquareFinl}) not only illustrates the algebraic decay of $\langle\Delta^2\rangle$ versus $L$ in large $L$ regime, but also explains its severe finite-size effect which originates from the subleading correction term. This result can serve as the mathematical explanation for the observation in our AFQMC simulations.

In Eq.~(\ref{eq:DeltaSquareFinl}), the constant $C_0$, which arises from short-range pairing correlations, should converge rapidly as the system size increases. This suggests that the full scaling form $\langle\Delta^2\rangle = aL^{-\eta}+bL^{-2}$, including the subleading term, can reliably describe the data down to relatively small system sizes. To verify this, we fit the numerical data of the rescaled pairing correlator $\langle\Delta^2\rangle\times L^{1/4}$ to the $(aL^{1/4-\eta}+bL^{-7/4})$ form for $U/t=-4,\mu/t=0.25$. As shown in Fig.~\ref{fig:FigOrderSquare}, this form works well for $L\ge12$, and yields $\eta$ values consistent with the asymptotic scaling $\langle\Delta^2\rangle\propto L^{-\eta}$ at $L\ge 32$ (see Fig.~1). This corroborates our finite-size analysis for $\langle\Delta^2\rangle$, particularly the result in Eq.~(\ref{eq:DeltaSquareFinl}).

\subsection{E. AFQMC simulation setup}
\label{sec:QMCSetUp}

Our calculations have reached $L=64$ system with lowest temperature $T/t=0.05$, which is significantly larger than ever before. We adopt $\Delta\tau t=0.04$ for $U/t=-4$, $\Delta\tau t=0.035$ for $U/t=-6$ and $\Delta\tau t=0.02$ for $U/t=-10$, for which the Trotter error is eliminated from varying $\Delta\tau t$ tests. We apply periodic boundary conditions in both directions for all the calculations. As for the statistics, we typically ensure that the relative errors of condensate fraction and the pairing correlator fall in between $0.1\%$ and $1\%$.

\section{III. Supplementary results for fixed \texorpdfstring{$\mu/t$}{} calculations}
\label{sec:FixedMuQMC}

In this section, we first show the illustrative results of condensate fraction and pair wave function in Sec. III A, and then present the additional results from AFQMC simulations with fixed $\mu/t$ for the main text, including $U/t=-4$ with $\mu=0.25,0.60,1.25$ and $U/t=-6$ with $\mu=0.5$ in Sec. III B, Sec. III C, Sec. III D, and Sec. III E. For such calculations, the fermion filling changes with both temperature and system size and it reaches converged value with large enough system size. As a summary, we present all the results of fermion filling and double occupancy for the above simulations in Sec. III F.

\subsection{A. Illustrative results of condensate fraction and pair wave function}
\label{sec:Demonstrate}

\begin{figure}[h]
\centering
\includegraphics[width=0.99\columnwidth]{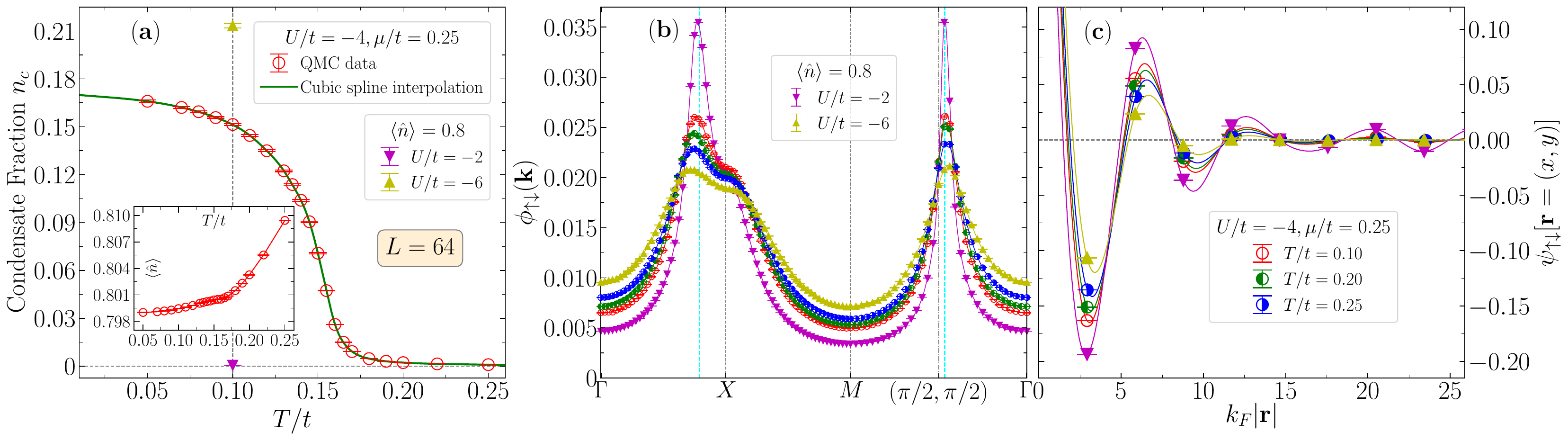}
\caption{\label{fig:Fig02CondPairWF} Demonstration of condensate fraction and pair wave function of 2D attractive Hubbard model for a $64\times 64$ system. Panel (a) plots high-precision condensate fraction results for $U/t=-4,\mu/t=0.25$ versus temperature, and also for $U/t=-2$ and $U/t=-6$ at $T/t=0.10$ with fermion filling $\langle\hat{n}\rangle=0.8$ (the same filling as $U/t=-4$ at $T/t=0.1$ as indicated by the vertical dashed line). The inset shows the corresponding fermion filling $\langle\hat{n}\rangle$ for $U/t=-4$. Panel (b) illustrates the momentum-space pair wave function $\phi_{\uparrow\downarrow}(\mathbf{k})$ along the high symmetry line of $\boldsymbol{\Gamma}\to\mathbf{X}\to\mathbf{M}\to\boldsymbol{\Gamma}$ with $\boldsymbol{\Gamma}=(0,0)$, $\mathbf{X}=(\pi,0)$ and $\mathbf{M}=(\pi,\pi)$, while panel (c) shows the corresponding real-space pair wave function $\phi_{\uparrow\downarrow}(\mathbf{r})$ along the diagonal line ($x=y$) versus $k_F|\mathbf{r}|$. The noninteracting Fermi surface at $T=0$ determined from the filling $\langle\hat{n}\rangle=0.8$ is plotted as vertical cyan dahsed lines in (b), and the one in $\mathbf{M}\to\boldsymbol{\Gamma}$ segment defines $k_F$ used in the $x$-axis label in (c). Panels (b) and (c) share the same plot legends.}
\end{figure}

We first take $64\times 64$ system with $U/t=-4,\mu/t=0.25$ as an example to illustrate the high-precision AFQMC results of condensate fraction and pair wave function, as shown in Fig.~\ref{fig:Fig02CondPairWF}. 

The vanishing condensate fraction $n_c$ at high temperature regime $T/t>0.20$ indicates the normal phase. With decreasing temperature, the condensate fraction $n_c$ plotted in Fig.~\ref{fig:Fig02CondPairWF}(a) monotonically grows and develops an inflection point, which can be taken as a signature of entering the superfluid phase. As a comparison, we have also included the $n_c$ results for $U/t=-2$ and $U/t=-6$ at $T/t=0.10$ for the same filling $\langle\hat{n}\rangle=0.8$ as $U/t=-4$ at the same temperature. These results reveal increasing $n_c$ from weak to strong interactions. 

The pair wave function in both momentum-space as $\phi_{\uparrow\downarrow}(\bf{k})$ and real-space as $\psi_{\uparrow\downarrow}(\bf{r})$ for specific parameters are presented in Fig.~\ref{fig:Fig02CondPairWF}(b) and (c), respectively. The $\phi_{\uparrow\downarrow}(\bf{k})$ results have prominent peaks around Fermi surfaces, which numerically validates the basic picture of BCS theory stating that the pairing is mainly contributed by the fermions around Fermi surfaces for weakly interacting regime. The corresponding $\psi_{\uparrow\downarrow}(\bf{r})$ manifests dominating contribution from local pairing around $|\mathbf{r}|\sim0$, and then exhibits a wave structure in extended regime with periodicity $\lambda\sim 2\pi/k_F$ which is also consistent with BCS theory. The $\psi_{\uparrow\downarrow}(\bf{r})$ results also clearly demonstrates the contribution of nonlocal fermion pairing which is completely neglected in the definition of the pairing correlator $\langle\Delta^2\rangle$. These features of $\phi_{\uparrow\downarrow}(\bf{k})$ and $\psi_{\uparrow\downarrow}(\bf{r})$ become more significant for lower temperature (as shown for $U/t=-4$) and weaker interaction (such as $U/t=-2$ in spite of the tiny $n_c<10^{-3}$). With increasing $|U|/t$, the pairing is more extended in momentum space as $\phi_{\uparrow\downarrow}(\mathbf{k}=\mathbf{k}_F)$ is surpressed and scattered to other momentum, and in real space the Cooper pairs gradually evolves into tightly bounded molecules as $\psi_{\uparrow\downarrow}(\bf{r})$ shrinks to $|\mathbf{r}|\sim0$. This is the BCS-BEC crossover which has been extensively studied in interacting Fermi gas with contact attraction~\cite{Shihao2015,YuanYao2022}. 

These illustrative results in Fig.~\ref{fig:Fig02CondPairWF} explicitly shows that the combination of condensate fraction and pair wave functions captures the full pairing structure in the attracive fermion systems. 

\subsection{B. \texorpdfstring{$U/t=-4,\mu/t=0.25$}{}}
\label{sec:Um04Mu025}

Most of the numerical results for this parameter set is already presented in the main text. Here we present more results of comparisons between the condensate fraction $n_c$ and on-site pairing correlator $\langle\Delta^2\rangle$. 

\begin{figure}[ht!]
\centering
\includegraphics[width=0.98\columnwidth]{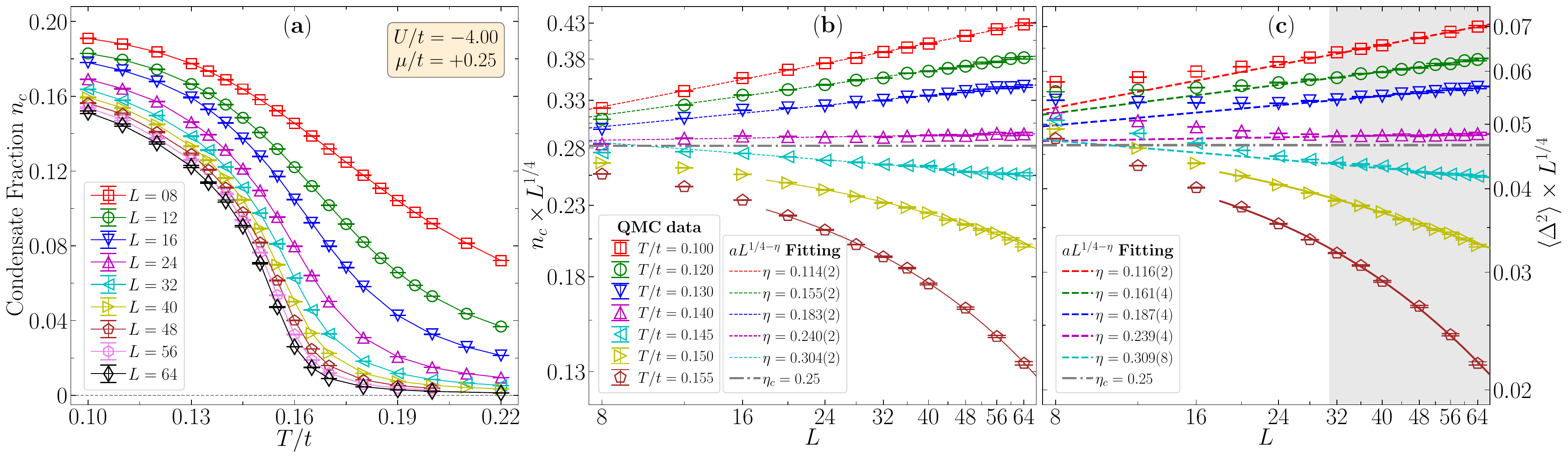}
\caption{\label{fig:Fig03CondnstU04Mu025} AFQMC results of condensate fraction $n_c$, and the rescaled results of $n_c$ and pairing correlator $\langle\Delta^2\rangle$ as $n_c\times L^{1/4}$ and $\langle\Delta^2\rangle\times L^{1/4}$ versus the linear system size $L$ in log-log plot, for $U/t=-4,\mu/t=0.25$. The dash lines in (b) and (c) are power-law fittings (for $T/t\le0.145$) with the solid lines as exponential fittings (for $T/t\ge0.150$), while the gray horizontal dash-dot lines represent $\eta_c=1/4$ seperating the superfluid and normal phases. The light gray shaded region in panel (b) marks the scaling regime for $\langle\Delta^2\rangle$ (as $L\ge 32$). Note that the algebraic scaling for $T/t=0.145$ should gradually crossovers to the exponential decay with larger system size that is still not accessable in our AFQMC calculations. }
\end{figure}
\begin{figure}[ht!]
\centering
\includegraphics[width=0.92\columnwidth]{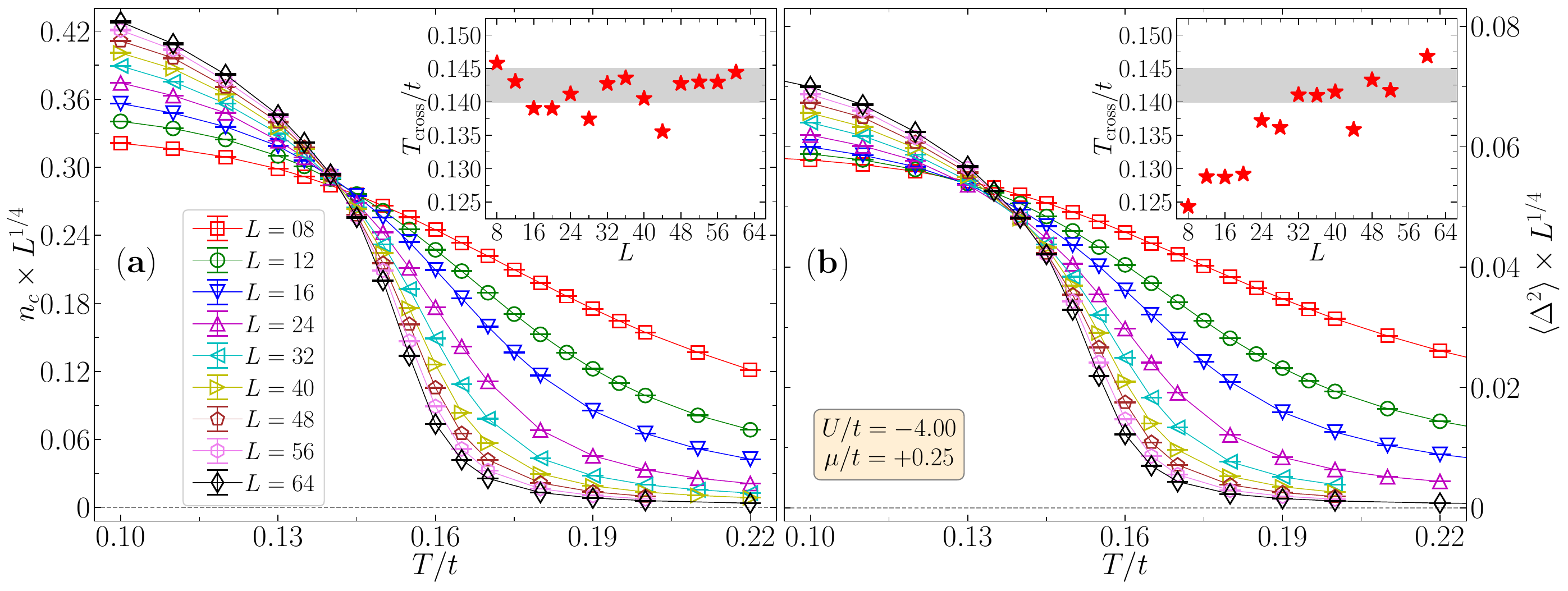}
\caption{\label{fig:Fig04CrossU04Mu025} Rescaled results of $n_c\times L^{1/4}$ and $\langle\Delta^2\rangle\times L^{1/4}$ versus temperature, for $U/t=-4,\mu/t=0.25$. Along with increasing $L$, the cross points between results from different system sizes should converge to the thermodynamic result of BKT transition temperature. The insets plot the cross points of results from $L$ and $L+4$, with the shading band indicating the bounded range of $T_{\rm BKT}/t$ obtained from the size scaling in Fig.~\ref{fig:Fig03CondnstU04Mu025}. }
\end{figure}

In Fig.~\ref{fig:Fig03CondnstU04Mu025}, we present the rescaled results of condensate fraction and pairing correlator as $n_c\times L^{1/4}$ and $\langle\Delta^2\rangle\times L^{1/4}$ versus the linear system size $L$ in the log-log plot, with the power-law fittings for $T/t\le0.145$ and exponential fittings for $T/t\ge0.150$. The gray horizontal dash-dot lines with $\eta_c=1/4$ seperates the superfluid and normal phases. It is clear that the condensate fraction results within $L\le20$ present the same bounded range of $T_{\rm BKT}$ as that considering the large system sizes. The $\langle\Delta^2\rangle$ results instead renders $0.12<T_{\rm BKT}<0.13$ within $L\le20$.

In Fig.~\ref{fig:Fig04CrossU04Mu025}, we plot the rescaled results of $n_c\times L^{1/4}$ and $\langle\Delta^2\rangle\times L^{1/4}$ versus temperature. From the finite-size scaling theory, the cross points between such results from different system sizes should converge to $T_{\rm BKT}$ along with increasing $L$. As ploted in the inset for condensate fraction, the cross points of results from $L$ and $L+4$ has little variation and are quite consistent with the bounded range ($0.140<T_{\rm BKT}<0.150$). As a comparison, the cross point of $\langle\Delta^2\rangle$ converges very slowly and it is obviously smaller than the results of $L\ge32$. These results again show the significant finite-size effect in the pairing correlator. 

In the inset of Fig.1 (a) in main text, we present the scaling of the exponent $\eta(T)$ to extract the BKT transition temperature. As mentioned in the main text, this quantity is equal to $\eta_c=1/4$ at the transition point and it is smaller than $1/4$ in the superfluid phase. We find that as $T\to T_{\rm BKT}^{-}$ this exponent satisfy a linear scaling with temperature. Such an extrapolation to $\eta(T)=\eta_c$ results in the transition temperature $T_{\rm BKT}/t=0.1420(7)$, consistent with the data collapse and logarithmic scaling of finite-size BKT transition temperature. 

\subsection{C. \texorpdfstring{$U/t=-6,\mu/t=0.50$}{}}
\label{sec:Um06Mu050}

Here we present similar results for $U/t=-6,\mu/t=0.50$, including temperature and size dependences of condensate fraction and pairing correlator, data collapse, and the energy results.

\begin{figure}[h]
\centering
\includegraphics[width=0.99\columnwidth]{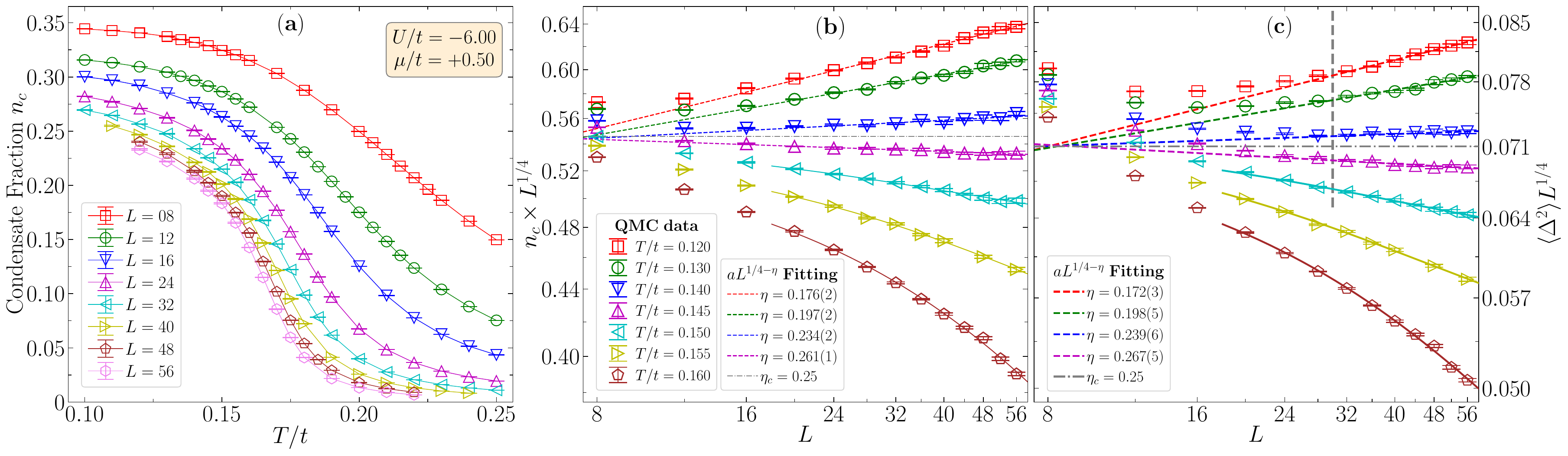}
\caption{\label{fig:Fig06CondnstU06Mu050} Condensate fraction $n_c$ and on-site pairing correlator $\langle\Delta^2\rangle$ results, for $U/t=-6,\mu/t=0.50$. Panel (a) plot the results of $n_c$ versus $T/t$. (b) and (c) show rescaled results of $n_c\times L^{1/4}$ and $\langle\Delta^2\rangle\times L^{1/4}$ versus the linear system size $L$ in log-log plot. The dashed lines are the power-law fitting of the QMC data with $L\ge20$ in (b), and with $L\ge32$ in (c), while the solid lines are the exponential fitting. Consistent exponents $\eta$ are obtained from these two quantites for $T/t\le0.145$. The gray, dash-dot lines highlight the exponent $\eta_c=1/4$ for the BKT transition, with the transition temperature bounded as $0.140<T_{\rm BKT}/t<0.145$ (with fermion filling $\sim$$0.5766$). The gray, vertical dash line in (c) marks the scaling regime of $L$ for $\langle\Delta^2\rangle$. }
\end{figure}

Figure~\ref{fig:Fig06CondnstU06Mu050} plots all the results of condensate fraction $n_c$ and pairing correlator $\langle\Delta^2\rangle$ for $U/t=-6,\mu/t=0.50$. The monotonically increasing $n_c$ with lowering temperature also shows an inflection point in the mediate temperature. The rescaled results of $n_c\times L^{1/4}$ and $\langle\Delta^2\rangle\times L^{1/4}$ versus $L$ in Fig.~\ref{fig:Fig06CondnstU06Mu050}(b) and (c) indicate the bounded range of transition temperature as $0.140<T_{\rm BKT}/t<0.145$. Consistent results of the exponent $\eta(T)$ extracted from both quantities again confirms the same size scaling behaviors of $n_c$ and $\langle\Delta^2\rangle$. Similar to the $U/t=-4$ case in Sec. III B, the strong finite-size effect in $\langle\Delta^2\rangle$ is pretty clear as shown in Fig.~\ref{fig:Fig06CondnstU06Mu050}(c), for which the $L<32$ results obviously deviate from the scaling behavior of $L\ge32$ results. If the simulations are limited to $L\le20$ and one only computes $\langle\Delta^2\rangle$, the result of $0.12<T_{\rm BKT}/t<0.13$ should be obtained. Instead, the condensate fraction shows much better scaling behavior and $L\le20$ results can already reach the correct bounded range of $T_{\rm BKT}/t$. A further $\eta(T)$ scaling, similar to that in the inset of Fig.1 (a) in main text, presents the transition temperature as $T_{\rm BKT}/t=0.1430(08)$.

\begin{figure}[ht!]
\centering
\includegraphics[width=0.92\columnwidth]{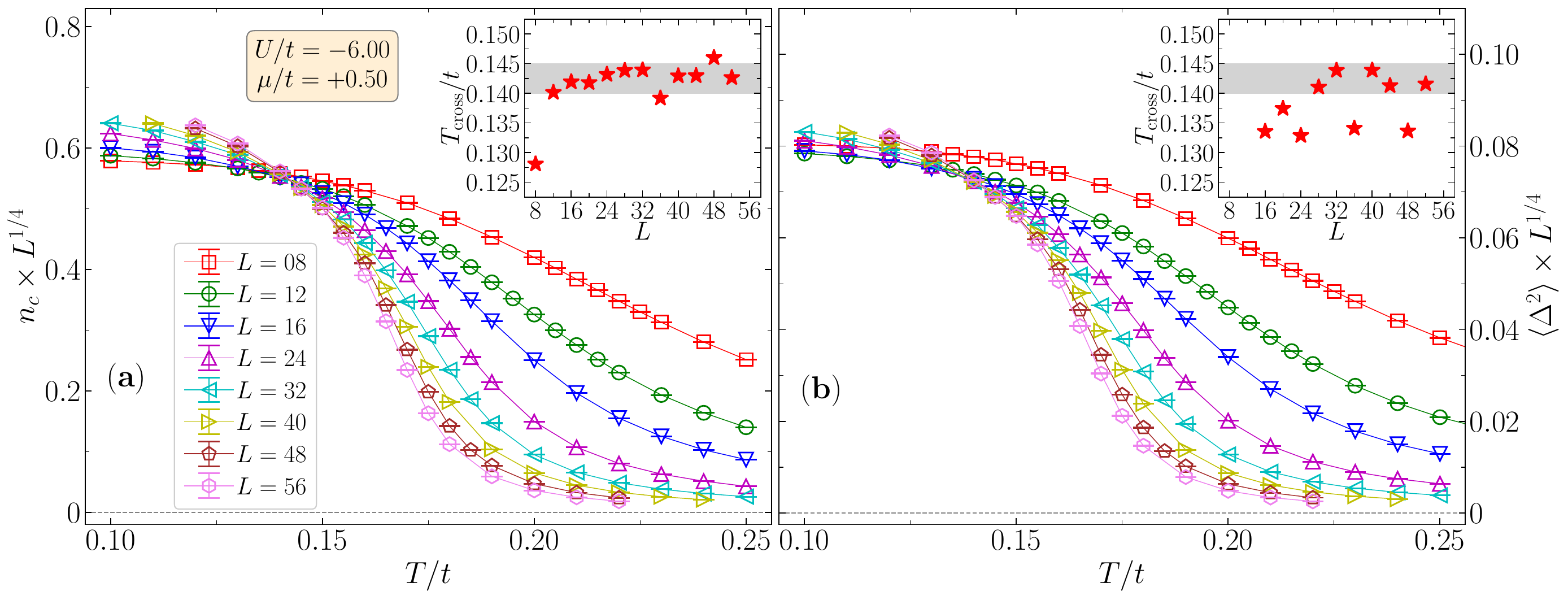}
\caption{\label{fig:Fig07CrossU06Mu050} Rescaled results of $n_c\times L^{1/4}$ and $\langle\Delta^2\rangle\times L^{1/4}$ versus temperature, for $U/t=-6,\mu/t=0.50$. The insets plot the cross points for results from $L$ and $L+4$, with the shading band indicating the bounded range of $T_{\rm BKT}/t$ obtained from the size scaling in Fig.~\ref{fig:Fig06CondnstU06Mu050}. }
\end{figure}
\begin{figure}[ht!]
\centering
\includegraphics[width=0.92\columnwidth]{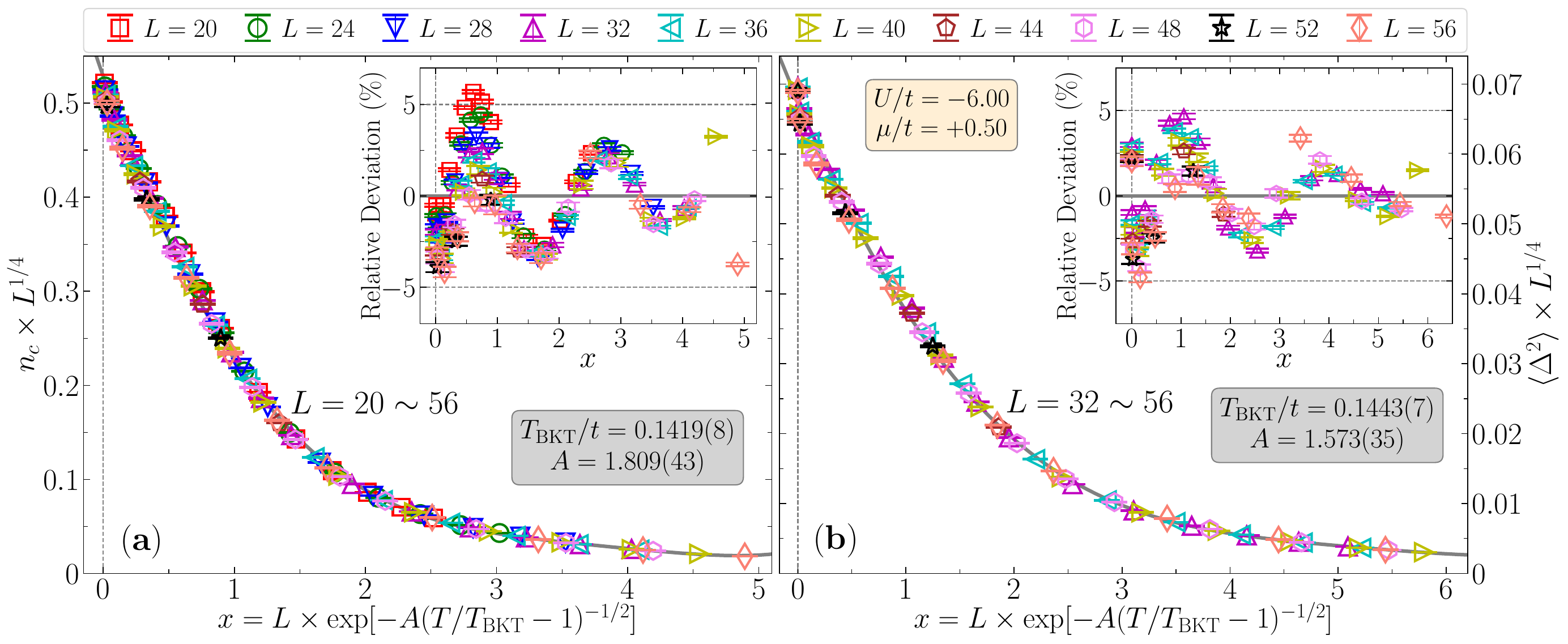}
\caption{\label{fig:Fig08CollapseU06Mu050} Data collapse for (a) condensate fraction $n_c$, and (b) on-site pairing correlator $\langle\Delta^2\rangle$ with fixed exponent $\eta_c=1/4$, for $U/t=-6,\mu/t=0.50$. The fitting process involves AFQMC data of $L=20$$\sim$$56$ for $n_c$ and of $L=32$$\sim$$56$ for $\langle\Delta^2\rangle$, respectively. The gray solid lines are the scaling invariant function $f(x)$, for which we adopt polynomials in $x$ for the fitting. The two insets sharing same legends plot the relative deviations from $f(x)$ of the fitting data. }
\end{figure}

\begin{figure}[ht!]
\centering
\includegraphics[width=0.92\columnwidth]{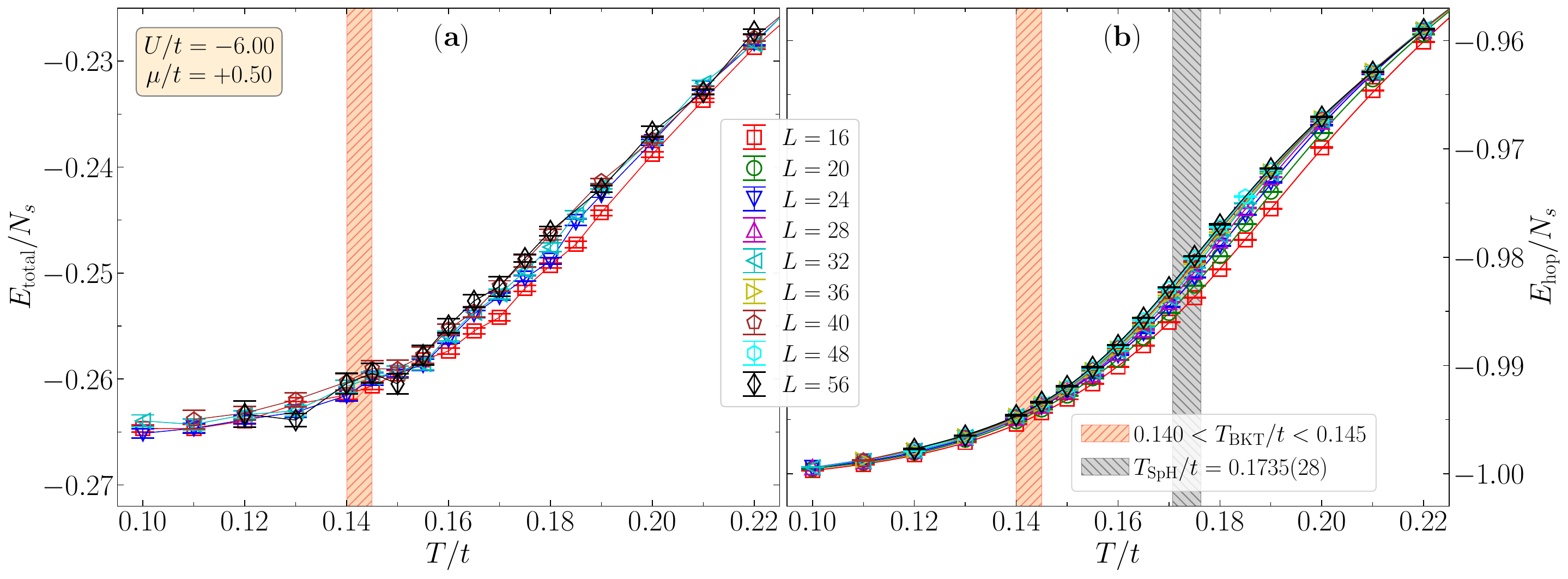}
\caption{\label{fig:Fig09EnergyU06Mu050} The results of (a) total energy per site $E_{\rm total}/N_s$ and (b) kinetic energy per site $E_{\rm hop}/N_s$ for $U/t=-6,\mu/t=0.50$. The bounded range of $T_{\rm BKT}/t$ from Fig.~\ref{fig:Fig06CondnstU06Mu050} and the peak location of $dE_{\rm hop}/dT$ (with error bar estimated by bootstrapping calculations) are indicated by the shading bands.}
\end{figure}

In Fig.~\ref{fig:Fig07CrossU06Mu050}, we plot the rescaled results of $n_c\times L^{1/4}$ and $\langle\Delta^2\rangle\times L^{1/4}$ versus temperature, with the cross points of results from $L$ and $L+4$ presented in the insets. Almost all of the crossing temperatures for condensate fraction fall into the correct bounded range of $T_{\rm BKT}/t$, which the fluctuation of cross points for pairing correlator at lower temperatures is prominent.

In Fig.~\ref{fig:Fig08CollapseU06Mu050}, we plot the data collapse results for $n_c$ and $\langle\Delta^2\rangle$. In the fitting process, we use the numerical data of $L=20\sim56$ for $n_c$ and of $L=32\sim56$ for $\langle\Delta^2\rangle$, respectively. These two quantities present the BKT transition temperature $T_{\rm BKT}=0.1419(8)$ and $0.1443(7)$, which are well consistent with the bounded range ($0.140<T_{\rm BKT}<0.150$) shown in Fig.~\ref{fig:Fig06CondnstU06Mu050} and the result from $\eta(T)$ scaling. The $n_c$ results clearly acquires a better data collapse quality than $\langle\Delta^2\rangle$.

In Fig.~\ref{fig:Fig09EnergyU06Mu050}, we show the results of total energy $E_{\rm total}$ and kinetic energy $E_{\rm hop}$ versus temperature for $U/t=-6,\mu/t=0.50$. For this case, the finite-size results of total energy are quite noisy as shown in Fig.~\ref{fig:Fig09EnergyU06Mu050}(a), prohibiting us from performing the extrapolations and obtaining precise results of specific heat. Instead, we find that kinetic energy also grows an inflection point (corresponding to the peak location of $dE_{\rm hop}/dT$) and its location explicitly converges with $L=40,48,56$ as $T_{\rm hop}/t=0.1735(28)$. For $U/t=-4,\mu/t=0.25$ case, we have verified that the peak location of $dE_{\rm hop}/dT$ and $dE_{\rm total}/dT$ are very close. So here we approximately take $T_{\rm hop}/t$ as the specific heat anomaly temperature as $T_{\rm SpH}/t=0.1735(28)$, which leads to a slight larger ratio of $T_{\rm SpH}/T_{\rm BKT}$$\sim$$1.2$ than that of $U/t=-4$ case.

\subsection{D. \texorpdfstring{$U/t=-4,\mu/t=0.60$}{}}
\label{sec:Um04Mu060}

Here we present similar results for $U/t=-4,\mu/t=0.60$, including temperature and size dependences of condensate fraction and pairing correlator and the specific heat results. Especially, for this case, we have also computed the superfluid density and determined $T_{\rm BKT}$ from this quantity. 

\begin{figure}[ht!]
\centering
\includegraphics[width=0.99\columnwidth]{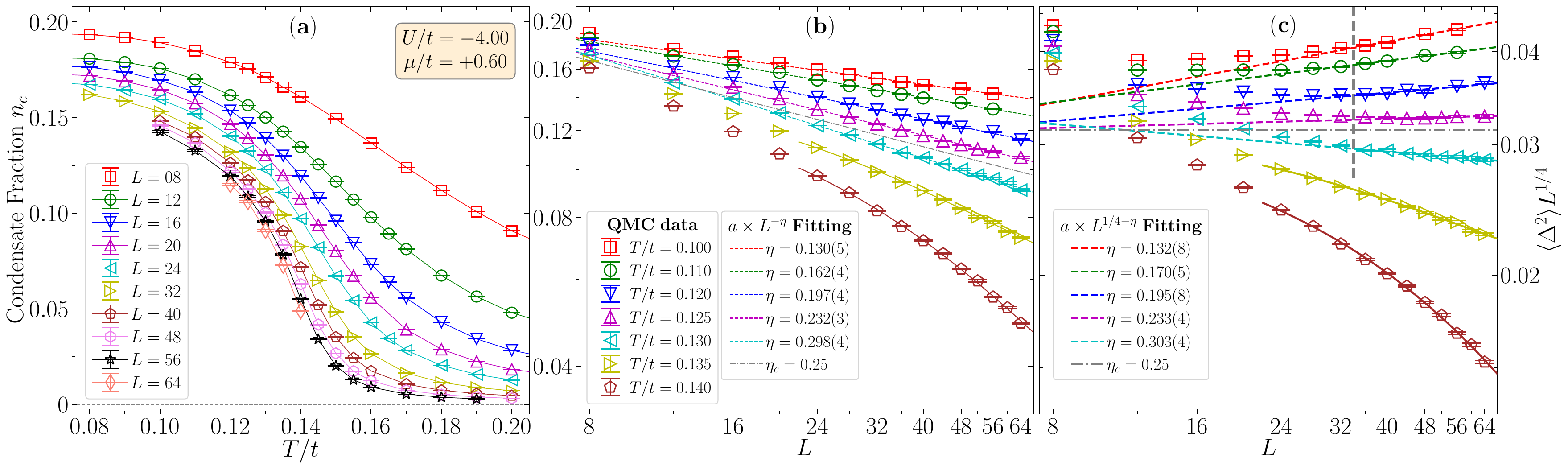}
\caption{\label{fig:Fig10CondnstU04Mu060} Condensate fraction $n_c$ and on-site pairing correlator $\langle\Delta^2\rangle$ results, for $U/t=-4,\mu/t=0.60$. Panel (a) and (b) plot high-precision results of $n_c$ versus $T/t$ and $L$, respectively; (c) plots $\langle\Delta^2\rangle\times L^{1/4}$ versus $L$. Panels (b) and (c) apply log-log plot, and share the same legends for QMC data. The dashed lines are the power-law fitting of the QMC data with $L\ge24$ in (b), and with $L\ge36$ in (c), while the solid lines are the exponential fitting. Consistent exponents $\eta$ are obtained from these two quantites for $T/t\le0.130$. The gray, dash-dot lines highlight the exponent $\eta_c=1/4$ for the BKT transition, with the transition temperature bounded as $0.125<T_{\rm BKT}/t<0.130$ (with fermion filling $\sim$$0.575$). The gray, vertical dash line in (c) marks the scaling regime of $L$ for $\langle\Delta^2\rangle$. }
\end{figure}

\begin{figure}[ht!]
\centering
\includegraphics[width=0.92\columnwidth]{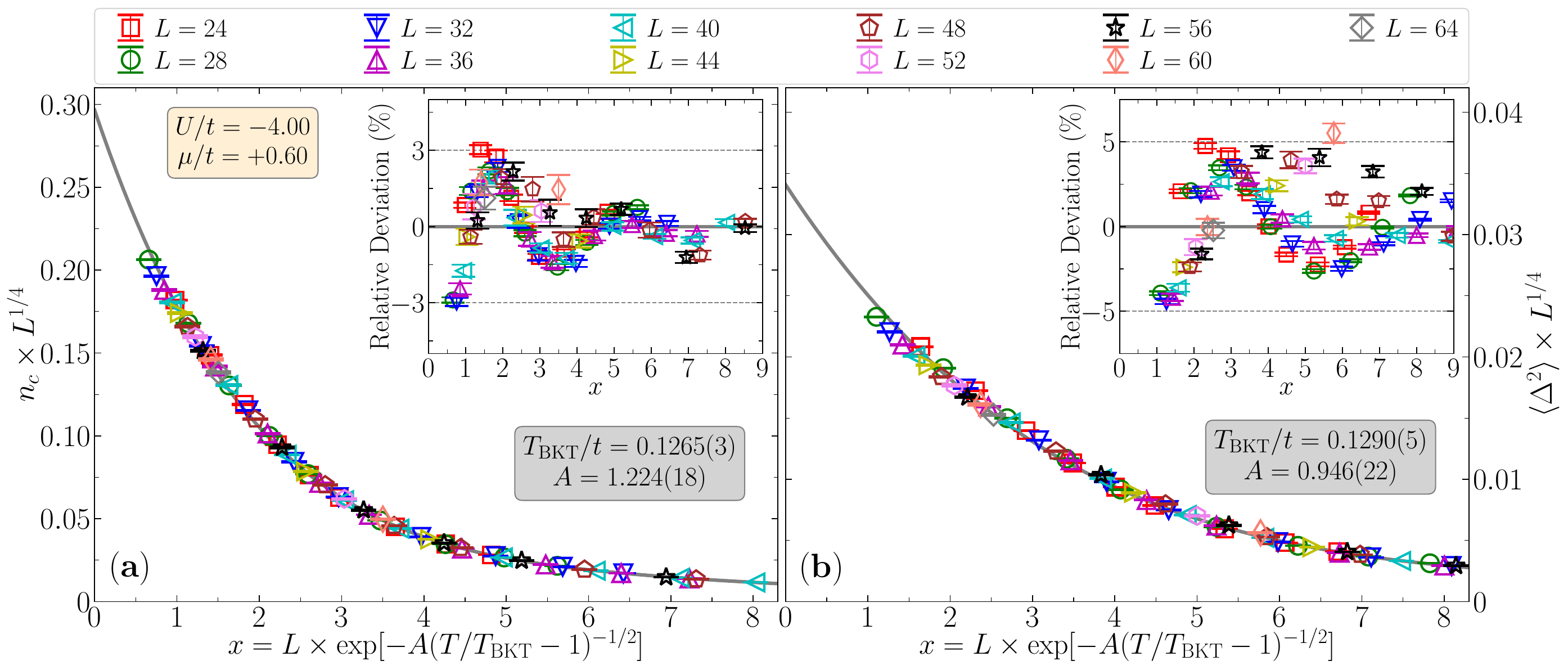}
\caption{\label{fig:CollapseU04Mu060} Data collapse for (a) condensate fraction $n_c$, and (b) on-site pairing correlator $\langle\Delta^2\rangle$ with fixed exponent $\eta_c=1/4$, for $U/t=-4,\mu/t=0.60$. The fitting process involves AFQMC data of $L=24$$\sim$$64$ for $n_c$ and $\langle\Delta^2\rangle$. The gray solid lines are the scaling invariant function $f(x)$, for which we adopt polynomials in $x$ for the fitting. The two insets sharing same legends plot the relative deviations from $f(x)$ of the fitting data. }
\end{figure}

\begin{figure}[ht!]
\centering
\includegraphics[width=0.92\columnwidth]{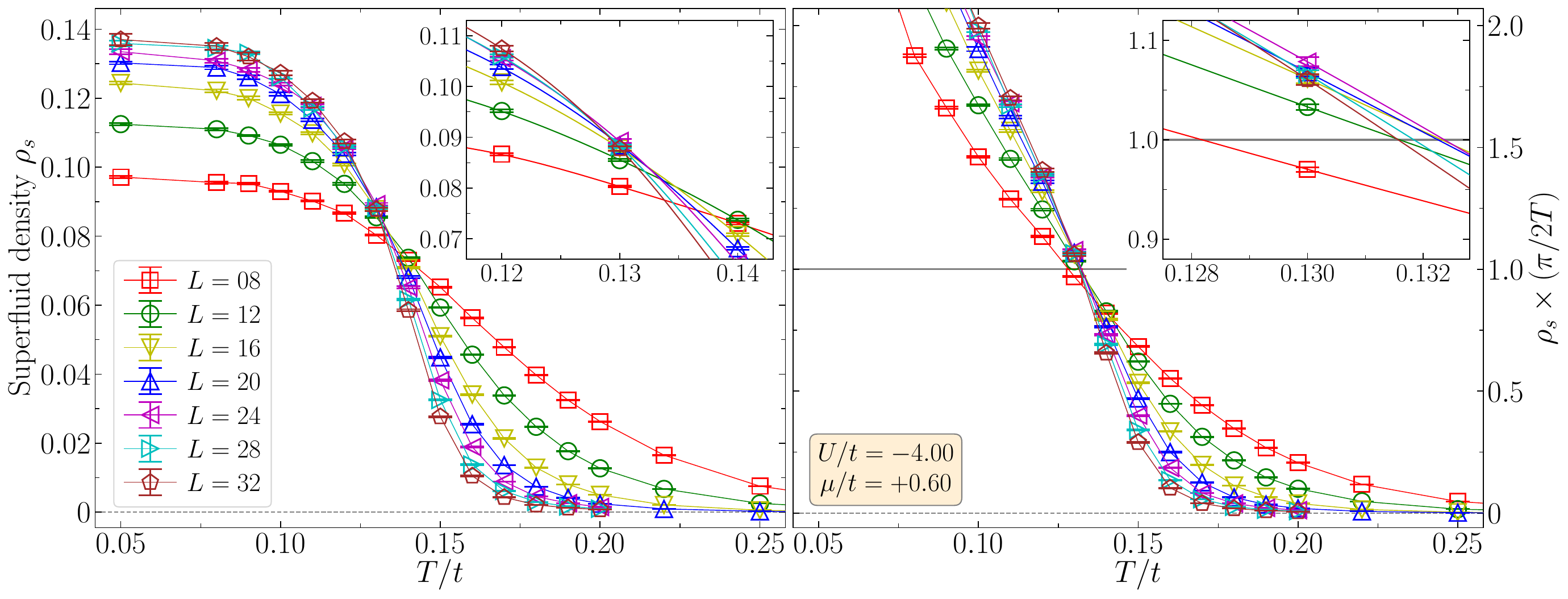}
\caption{\label{fig:Fig11SuperFluid} superfluid density $\rho_s$ results versus temperature for $L=8\sim32$ for $U/t=-4,\mu/t=0.60$. Panel (a) plots the bare $\rho_s$ results with the inset as zoom in of the data crossing region. Panel (b) plots $\rho_s\times(\pi/2T)$ versus temperature, with the inset showing the data curves crossing with unity. }
\end{figure}

\begin{figure}[ht!]
\centering
\includegraphics[width=0.70\columnwidth]{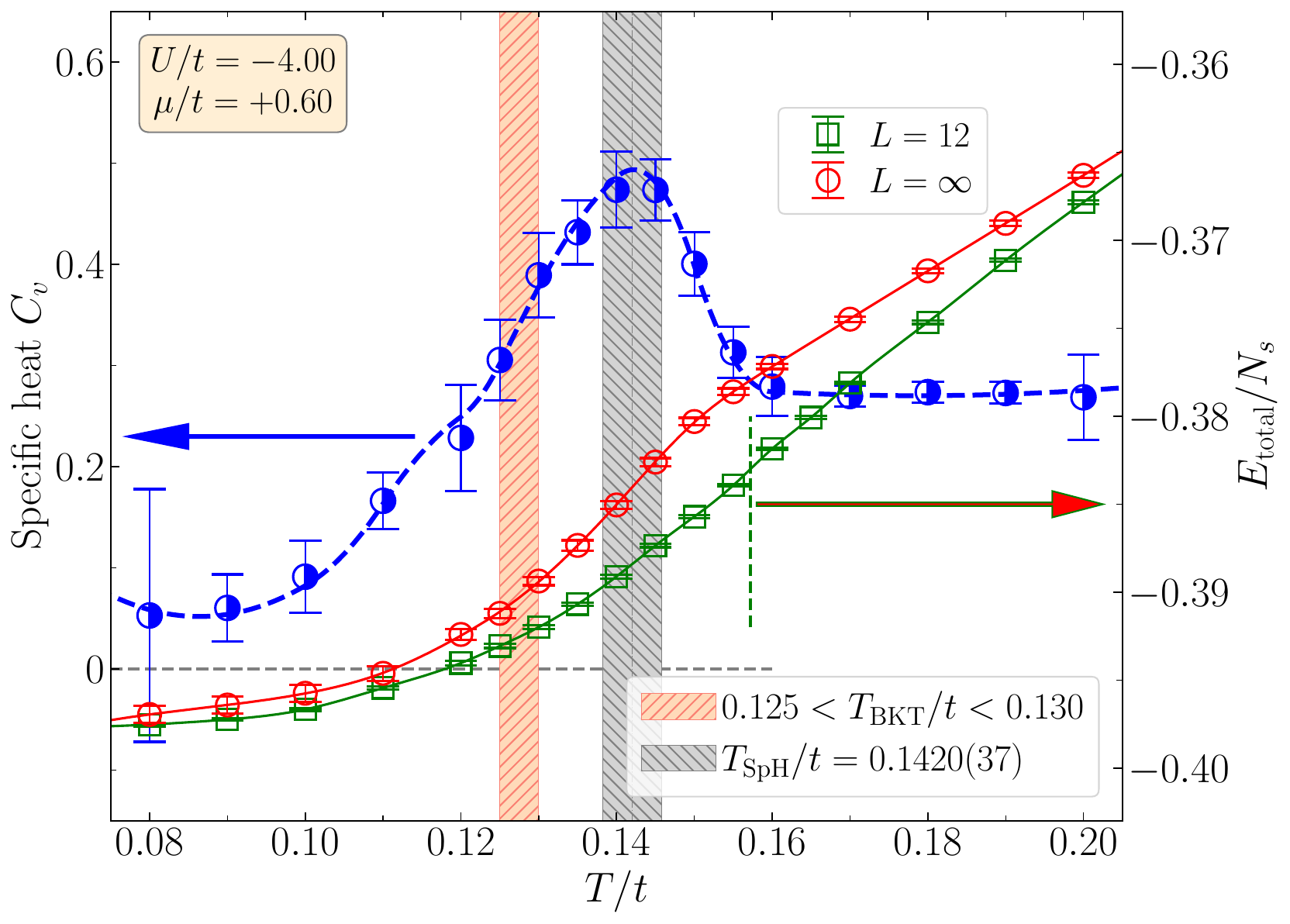}
\caption{\label{fig:Fig12Specific04Mu060} Specific heat $C_v$ and total energy per site $E/N_s$ versus $T/t$ for $U/t=-4,\mu/t=0.60$. Energy results of $L=12$ (green squares) and $L=\infty$ (red circles) with solid lines as cubic-spline fittings are plotted. TDL results for $C_v$ from the fitting curve of $L=\infty$ energies (blue dash line) and numerical differentiation using total-variation regularization (blue circles) are shown. The peak location of $dE/dT$ for $L=12$ is marked by vertical green dash line as a comparison. The bounded range of $T_{\rm BKT}/t$ and the peak location of specific heat $T_{\rm SpH}/t$ (with error bar estimated by bootstrapping calculations during fitting the energies) are indicated by the shading bands, leading to the anomaly as $T_{\rm SpH}/T_{\rm BKT}$$\sim$$1.10$.}
\end{figure}

Figure~\ref{fig:Fig10CondnstU04Mu060} plots all the results of condensate fraction $n_c$ and pairing correlator $\langle\Delta^2\rangle$ for $U/t=-4,\mu/t=0.60$. As shown in panel (b), the algebraic scaling in superfluid phase and exponential decaying in normal state of $n_c$ is clear, which results in the bounded range of the transition temperature as $0.125<T_{\rm BKT}/t<0.130$. Similar scalings and consistent results of $\eta(T)$ for the pairing correlator can only be reached for $L\ge36$ results, showing even more severe finite-size effect than that for $U/t=-4,\mu/t=0.25$ case. This is attributed to the smaller fermion filling of this case, for which the nonlocal fermion pairing have larger contributions and thus makes the finite-size effect in on-site pairing correlator more significant. If the simulations are limited to $L\le20$ and one only computes $\langle\Delta^2\rangle$, the result of $0.10<T_{\rm BKT}/t<0.11$ should be obtained. Instead, the condensate fraction can still reach the correct range of $T_{\rm BKT}/t$ within $L\le20$ results. A further $\eta(T)$ scaling presents the transition temperature as $T_{\rm BKT}/t=0.1264(17)$.

Figure~\ref{fig:CollapseU04Mu060} plots the data collapse results for $n_c$ and $\langle\Delta^2\rangle$ for $U/t=-4,\mu/t=0.60$. These two quantities present the BKT transition temperature $T_{\rm BKT}=0.1265(3)$ and $0.1290(5)$, which are consistent with the bounded range $0.125<T_{\rm BKT}<0.130$ shown in Fig.~\ref{fig:Fig10CondnstU04Mu060} and the result from $\eta(T)$ scaling. 

In Fig.~\ref{fig:Fig11SuperFluid}, we present the AFQMC results of superfluid density $\rho_s$ versus temperature for $U/t=-4,\mu/t=0.60$. According to BKT theory, this quantity should acquire a universal jump at the BKT transition point. However, due to the finite-size simulations, we can only obtain continuous results of $\rho_s$ across the transition, and we can see that the curve of $\rho_s$ becomes more and more sharp with increasing system size. For $L\ge24$, the AFQMC results of $\rho_s$ are quite noisy at low temperatures, but the results from $L=28$ and $L=32$ show convergence at $T/t\le0.10$. Besides, $\rho_s$ also reach convergence with temperature for $T/t\le0.08$, indicating that the superfluid density is about $0.14$ even for the ground state. 

There are several ways to extract the BKT transition temperature from superfluid density. First, the crossing points of $\rho_s$ from different system sizes as shwon in Fig.~\ref{fig:Fig11SuperFluid}(a) and its inset should converge to $T_{\rm BKT}/t$ towards the thermodynamic limit, resulting in the jump of $\rho_s$ and the transition point. We can see that for $L=24,28,32$, the crossing points almost converge and resides in the range of $(0.125,0.130)$, which is well consistent with that obtained from the size scaling of condensate fraction in Fig.~\ref{fig:Fig10CondnstU04Mu060}(b). Second, superfluid density at just below the transition temperature satisfies the relation $\rho_s(T\to T_{\rm BKT}^+)=2T_{\rm BKT}/\pi$. Based on this relation, we can take the crossing points of $\rho_s\times(\pi/2T)$ with unity as the finite-size transition temperature $T_{\rm BKT}(L)$~\cite{Filinov2010}, and this result should also satisfy the logarithmic correlation relation $T_{\rm BKT}(L)=T_{\rm BKT}(L=\infty)+a/(\ln bL)^2$. From the results shown in Fig.~\ref{fig:Fig11SuperFluid}(b), $T_{\rm BKT}(L)$ from such crossing analysis showes nonmotonic finite-size effect, as it first increases from $L=8$ to $L=20$ and then decreases with $L\ge24$. The crossing point of $L=32$ is $T_{\rm BKT}(L=32)$$\sim$$0.1315$. Performing AFQMC calculations of dynamic correlations for $L>32$ will surely consume much more computational resources. Considering the moving slow convergence of this crossing point with increasing system size, the final $T_{\rm BKT}(L=\infty)$ result from this method should also be consistent with the bounded range of $(0.125,0.130)$. 

In Fig.~\ref{fig:Fig12Specific04Mu060}, we show the results of specific heat $C_v$ and total energy per site $E/N_s$ for $U/t=-4,\mu/t=0.60$. A clear peak in results of specific heat appears at $T_{\rm SpH}/t=0.1420(37)$, above the BKT transition. The cubic-spline fitting process and numerical differentiation calculation for the total energy present consistent results for this specific heat anomaly as $T_{\rm SpH}/T_{\rm BKT}$$\sim$$1.12$. We also note that the height of the peak in specific heat here is smaller than that for $U/t=-4,\mu/t=0.25$.

\subsection{E. \texorpdfstring{$U/t=-4,\mu/t=1.25$}{}}
\label{sec:Um04Mu125}

Here we present similar results for $U/t=-4,\mu/t=1.25$, including temperature and size dependences of condensate fraction and pairing correlator and the specific heat results. 

\begin{figure}[ht!]
\centering
\includegraphics[width=0.99\columnwidth]{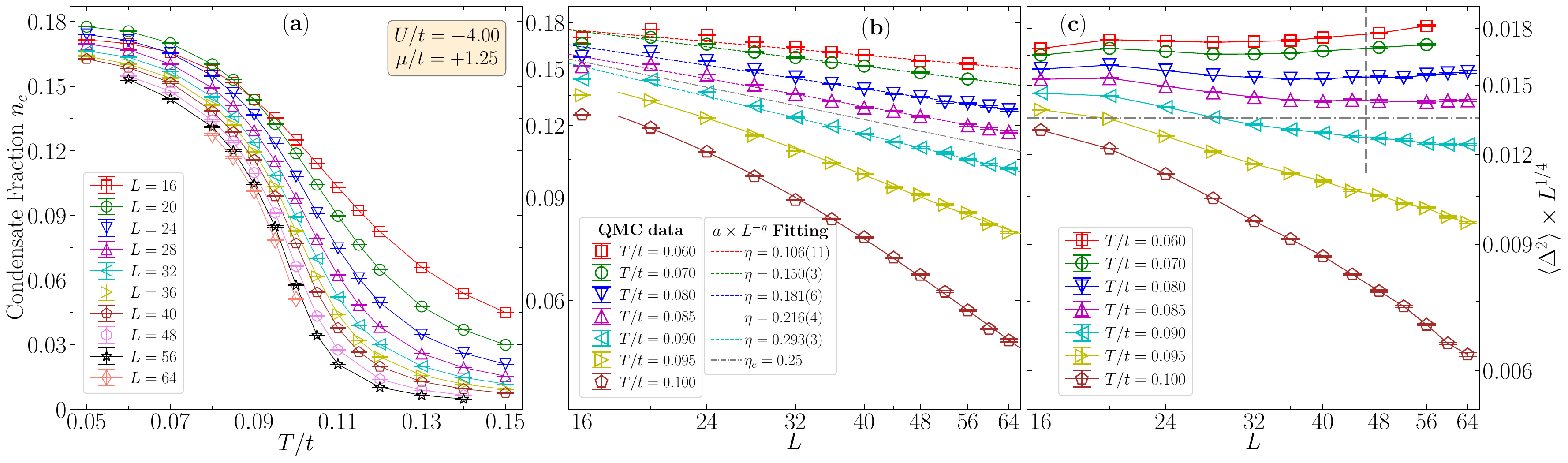}
\caption{\label{fig:Fig13CondnstU04Mu125} Condensate fraction $n_c$ and on-site pairing correlator $\langle\Delta^2\rangle$ results, for $U/t=-4,\mu/t=1.25$. Panel (a) and (b) plot high-precision results of $n_c$ versus $T/t$ and $L$, respectively; (c) plots $\langle\Delta^2\rangle\times L^{1/4}$ versus $L$. Panels (b) and (c) apply log-log plot. In (b), dashed lines are the power-law fitting of the QMC data, while solid lines are exponential fittings. The gray, dash-dot lines highlight the exponent $\eta_c=1/4$ for the BKT transition, with the transition temperature bounded as $0.085<T_{\rm BKT}/t<0.090$ (with fermion filling $\sim$$0.292$). The gray, vertical dash line in (c) marks the scaling regime of $L$ for $\langle\Delta^2\rangle$. }
\end{figure}

\begin{figure}[ht!]
\centering
\includegraphics[width=0.56\columnwidth]{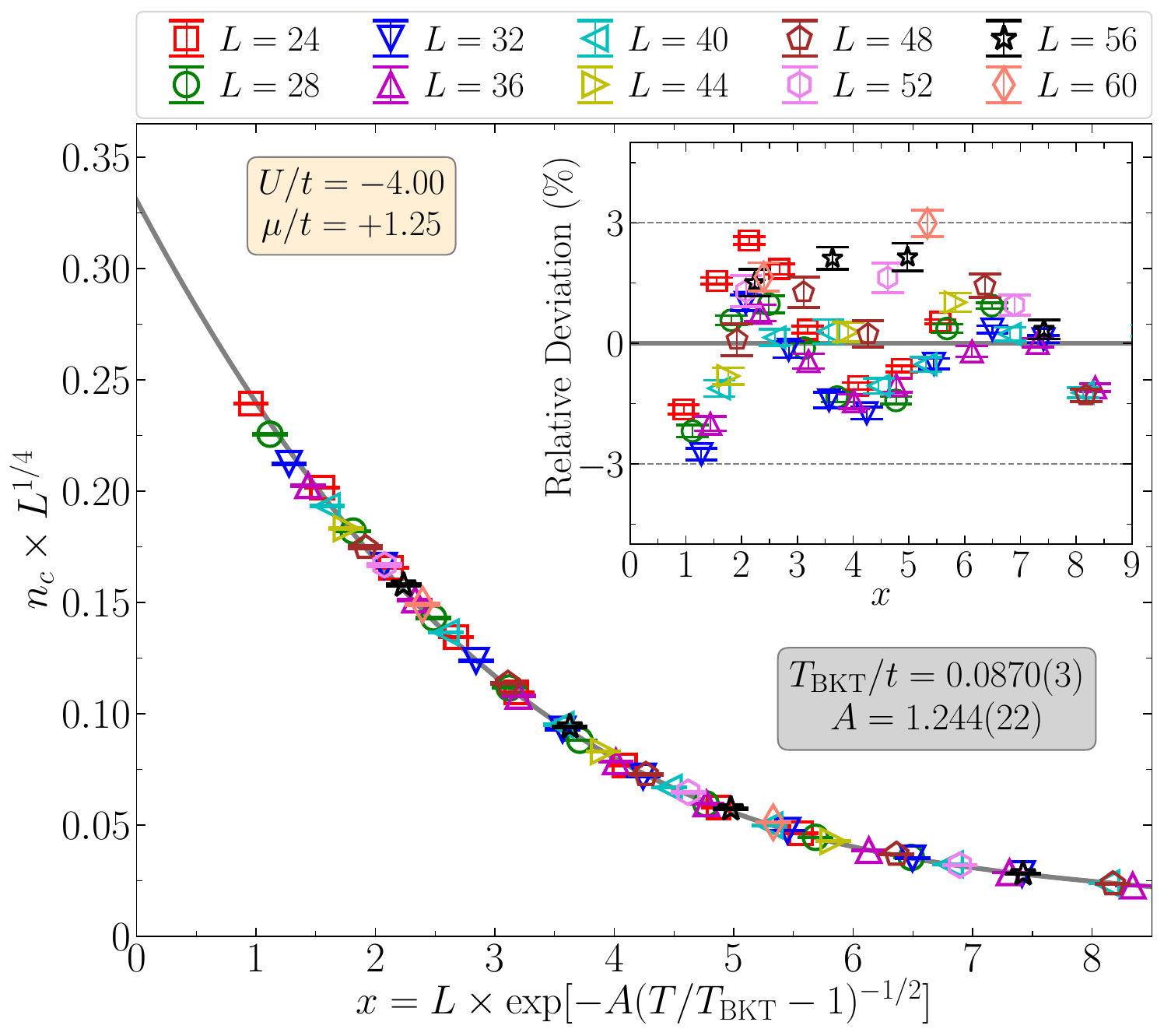}
\caption{\label{fig:CollapseU04Mu125} Data collapse for (a) condensate fraction $n_c$, and (b) on-site pairing correlator $\langle\Delta^2\rangle$ with fixed exponent $\eta_c=1/4$, for $U/t=-4,\mu/t=1.25$. The fitting process involves AFQMC data of $L=24$$\sim$$64$ for $n_c$ and $\langle\Delta^2\rangle$. The gray solid lines are the scaling invariant function $f(x)$, for which we adopt polynomials in $x$ for the fitting. The two insets sharing same legends plot the relative deviations from $f(x)$ of the fitting data. }
\end{figure}

\begin{figure}[ht!]
\centering
\includegraphics[width=0.63\columnwidth]{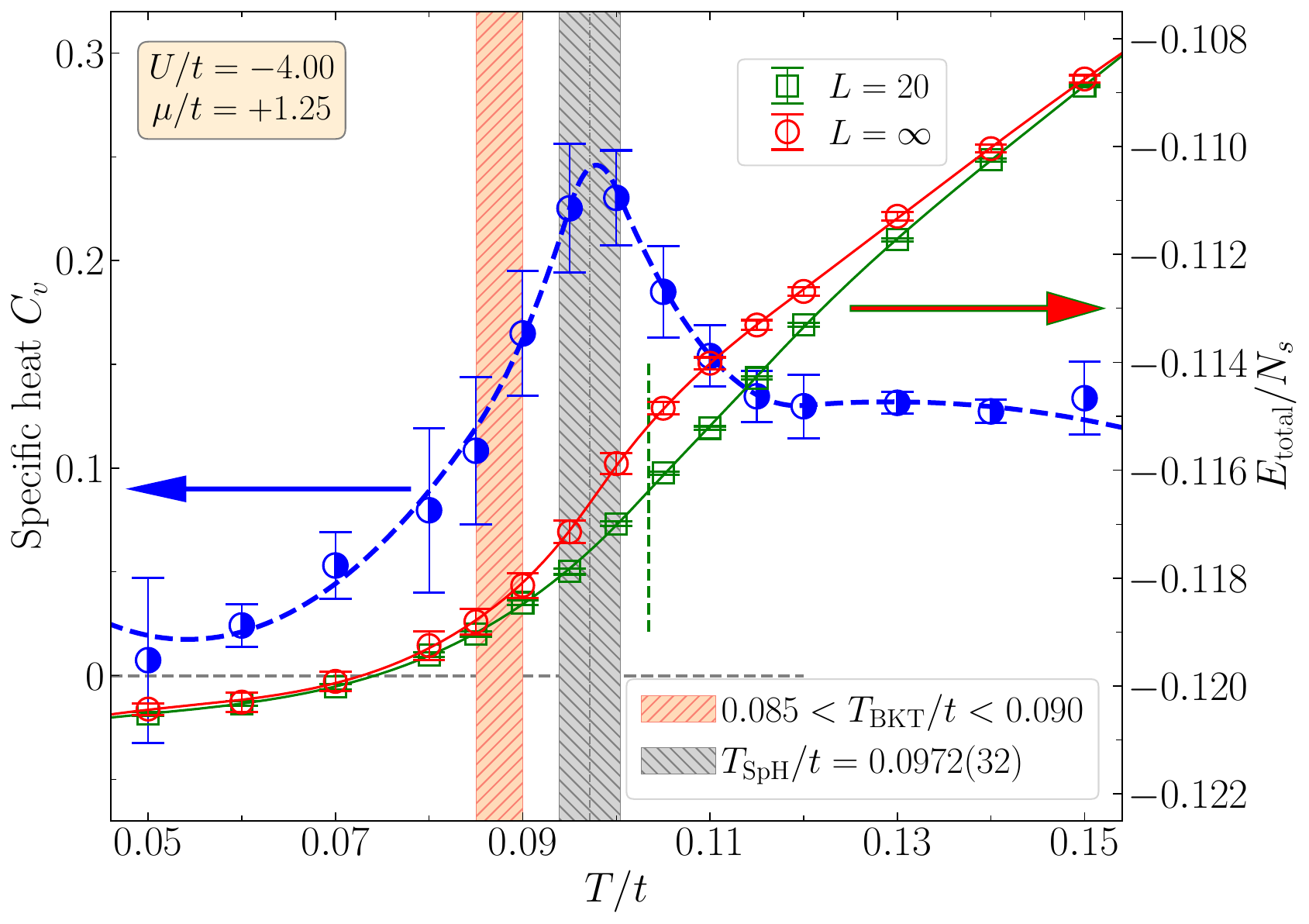}
\caption{\label{fig:Fig14Specific04Mu125} Specific heat $C_v$ and total energy per site $E/N_s$ versus $T/t$ for $U/t=-4,\mu/t=1.25$. Energy results of $L=20$ (green squares) and $L=\infty$ (red circles) with solid lines as cubic-spline fittings are plotted. TDL results for $C_v$ from the fitting curve of $L=\infty$ energies (blue dash line) and numerical differentiation using total-variation regularization (blue half-filled circles) are shown. The peak location of $dE/dT$ for $L=20$ is marked by vertical green dash line as a comparison. The bounded range of $T_{\rm BKT}/t$ and the peak location of specific heat $T_{\rm SpH}/t$ (with error bar estimated by bootstrapping calculations during fitting the energies) are indicated by the shading bands, leading to the anomaly as $T_{\rm SpH}/T_{\rm BKT}$$\sim$$1.10$. }
\end{figure}

Figure~\ref{fig:Fig13CondnstU04Mu125} plots all the results of condensate fraction $n_c$ and pairing correlator $\langle\Delta^2\rangle$ for $U/t=-4,\mu/t=1.25$. As shown in panel (b), the algebraic scaling in superfluid phase and exponential decaying in normal state of $n_c$ is clear, which results in the bounded range of the transition temperature as $0.085<T_{\rm BKT}/t<0.090$. On the other hand, the on-site pairing correlator gets too small for this case, due to the rather small filling $<0.30$. From Fig.~\ref{fig:Fig13CondnstU04Mu125}(c), we can see that $\langle\Delta^2\rangle\times L^{1/4}$ for $T/t=0.085$ reaches minimum at $L=40\sim48$ and almost reach plateau with even larger system size. This means that the $L=40$ results of $\langle\Delta^2\rangle$ still suffer strong finite-size effect. For this low filling case, the condensate fraction also show some finite-size effect for $L\le28$, but it is still much better than that of $\langle\Delta^2\rangle$. A further $\eta(T)$ scaling presents the transition temperature as $T_{\rm BKT}/t=0.0872(15)$.

Figure~\ref{fig:CollapseU04Mu125} plots the data collapse results for $n_c$ for $U/t=-4,\mu/t=1.25$. The fitting with least-squares criterion presents the transition temperature $T_{\rm BKT}=0.0870(3)$, which is consistent with the bounded range $0.085<T_{\rm BKT}<0.090$ shown in Fig.~\ref{fig:Fig13CondnstU04Mu125} and the result from $\eta(T)$ scaling. 

In Fig.~\ref{fig:Fig14Specific04Mu125}, we show the results of specific heat $C_v$ and total energy per site $E/N_s$ for $U/t=-4,\mu/t=1.25$. A clear peak in results of specific heat appears at $T_{\rm SpH}/t=0.0972(32)$, above the BKT transition. The cubic-spline fitting process and numerical differentiation calculation for the total energy present consistent results for this specific heat anomaly as $T_{\rm SpH}/T_{\rm BKT}$$\sim$$1.11$. We also note that the height of the peak in specific heat gradually decreases for $U/t=-4$ from $\mu/t=0.25$ to $\mu/t=0.60$ and to $\mu/t=1.25$ as towards the lower filling regime.

\subsection{F. Summary of all the AFQMC simulations}
\label{sec:SumOfAll}

Here we present a summary for all the AFQMC simulation results in Sec. II, and also present the results of fermion filling and double occupancy, which might be used as benchmark for future experiments, other theoretical studies and many-body calculations. 

\begin{figure}[ht!]
\centering
\includegraphics[width=0.99\columnwidth]{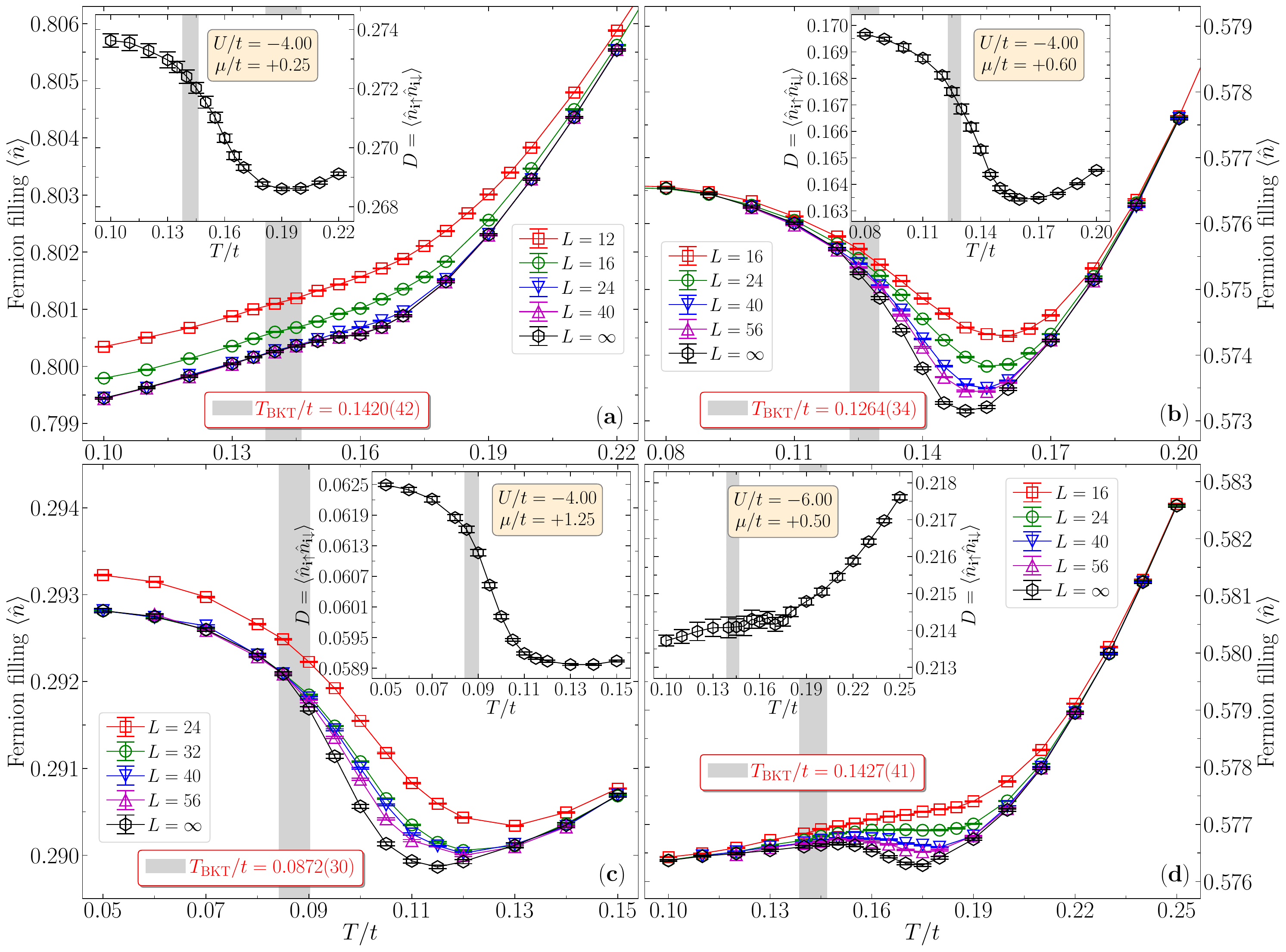}
\caption{\label{fig:Fig15SumOfAll} The fermion filling (the main plots) and double occupancy (the insets) versus temperature and system size for all the four sets of parameters in Sec. III B $\sim$ Sec. III E. For $U/t=-4,\mu/t=0.25$ and $U/t=-6,\mu/t=0.50$, the $T_{\rm BKT}$ results are the ones obtained from scaling of $T_{\rm BKT}(L)$ as shown in Fig.1 in the main text, while for the other two cases the results are extracted from $\eta(T)$ scaling. The corresponding BKT transition temperatures are listed and indicated by the vertical, gray bands. }
\end{figure}

We have mainly performed the AFQMC simulations for 2D attracive Hubbard model for $U/t=-4$ with fixed $\mu/t=0.25,0.60,1.25$ and $U/t=-6$ with fixed $\mu/t=0.50$. Our calculations reach the linear system size $L=64$ with the lowest temperature $T/t=0.05$. We mainly concentrate on the condensate fraction $n_c$ and on-site pairing correlator $\langle\Delta^2\rangle$ around the BKT transition. All of our numerical results reveal that these two quantities have the same size scaling behaviors, i.e., algebraic scaling ($\propto L^{-\eta}$) in superfluid phase and exponential decaying in normal state. We also find that the extensively studied pairing correlator suffers from severe finite-size effect (as deviation from the correct scaling in small sizes). As a comparison, the condensate fraction shows significantly better size scaling in small systems, rendering it as a more appropriate quantity to determine the BKT transition temperature. 

In Fig.~\ref{fig:Fig15SumOfAll}, we present the summarization results of fermion filling (main plots) and double occupancy (the inset) versus temperature and system size for all the four sets of parameters in Sec. III B $\sim$ Sec. III E. The $L=\infty$ results are obtained from finite-size extrapolations. The corresponding BKT transition temperatures are also presented in the plots.

\section{IV. Supplementary results for fixed fermion filling calculations}
\label{sec:FixedNtQMC}

In this section, we further present the results of AFQMC simulations with fixed fermion filling $\langle\hat{n}\rangle=0.5$. Results for two representative different interaction strengths $U/t=-4$ and $U/t=-10$ are obtained. These results also explicitly shows that condensate fraction and on-site pairing correlator have the same size scaling, with the finite-size effect in the former is significantly smaller than the latter. 

\subsection{A. \texorpdfstring{$U/t=-4,\langle\hat{n}\rangle=0.5$}{}}
\label{sec:Um04nT050}

\begin{figure}[ht!]
\centering
\includegraphics[width=0.99\columnwidth]{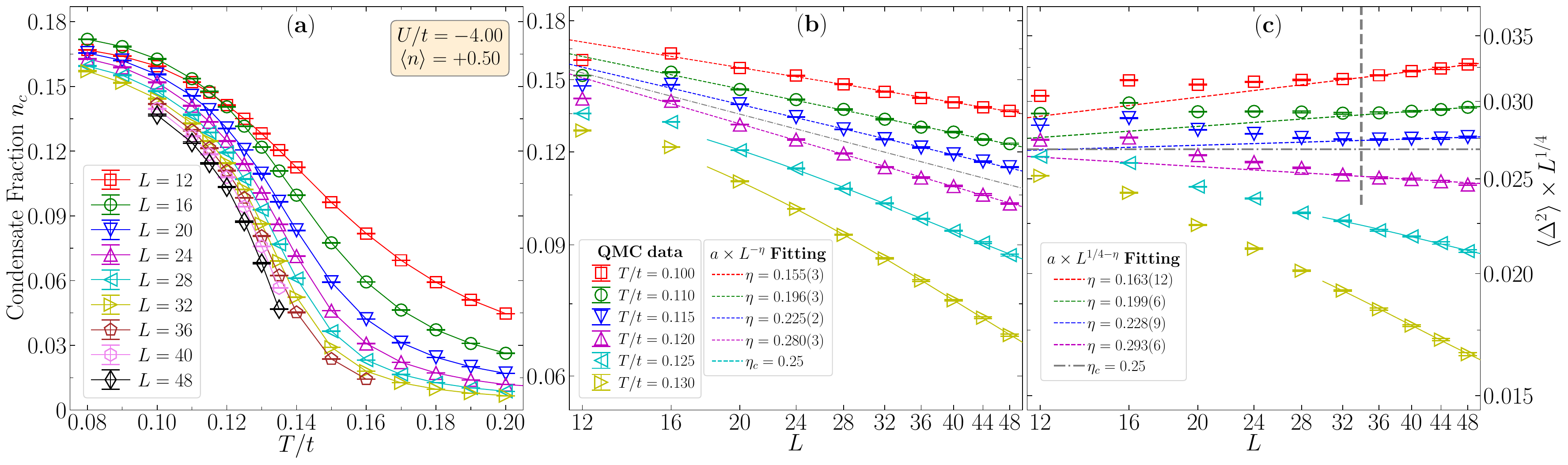}
\caption{\label{fig:Fig16CondnstU04nT050} Condensate fraction $n_c$ and on-site pairing correlator $\langle\Delta^2\rangle$ results, for $U/t=-4,\langle\hat{n}\rangle=0.50$. Panel (a) and (b) plot high-precision results of $n_c$ versus $T/t$ and $L$, respectively; (c) plots $\langle\Delta^2\rangle\times L^{1/4}$ versus $L$. Panels (b) and (c) apply log-log plot. The dashed lines (for $T/t\le0.120$) are the power-law fitting of the QMC data with $L\ge20$ in (b), and with $L\ge36$ in (c), while the solid lines (for $T/t\ge0.125$) are the exponential fitting. Consistent exponents $\eta$ are obtained from these two quantites for $T/t\le0.120$. The gray, dash-dot lines highlight the exponent $\eta_c=1/4$ for the BKT transition, with the transition temperature bounded as $0.115<T_{\rm BKT}/t<0.120$. The gray, vertical dash line in (c) marks the scaling regime as $L\ge36$ for $\langle\Delta^2\rangle$. }
\end{figure}

Figure~\ref{fig:Fig16CondnstU04nT050} plots all the results of condensate fraction $n_c$ and pairing correlator $\langle\Delta^2\rangle$ for $U/t=-4,\langle\hat{n}\rangle=0.50$. As shown in panel (b), the algebraic scaling in superfluid phase and exponential decaying in normal state of $n_c$ is clear, which results in the bounded range of the transition temperature as $0.115<T_{\rm BKT}/t<0.120$. On the other hand, the on-site pairing correlator suffers severe finite-size effect and only $L\ge36$ results can reach the correct algebraic scaling. If the simulation is limited to $L=20$, the result of $T_{\rm BKT}/t\sim 0.10$ should be obtained. Instead, the condensate fraction results can correctly present the transition temperature within $L\le20$. A further scaling for the exponent $\eta(T)$ (using the $\eta$ results of $T/t=0.110$ and $0.115$) determines the transition temperature as $T_{\rm BKT}/t=0.1193(19)$.

\subsection{B. \texorpdfstring{$U/t=-10,\langle\hat{n}\rangle=0.5$}{}}
\label{sec:Um10nT050}

\begin{figure}[ht!]
\centering
\includegraphics[width=0.99\columnwidth]{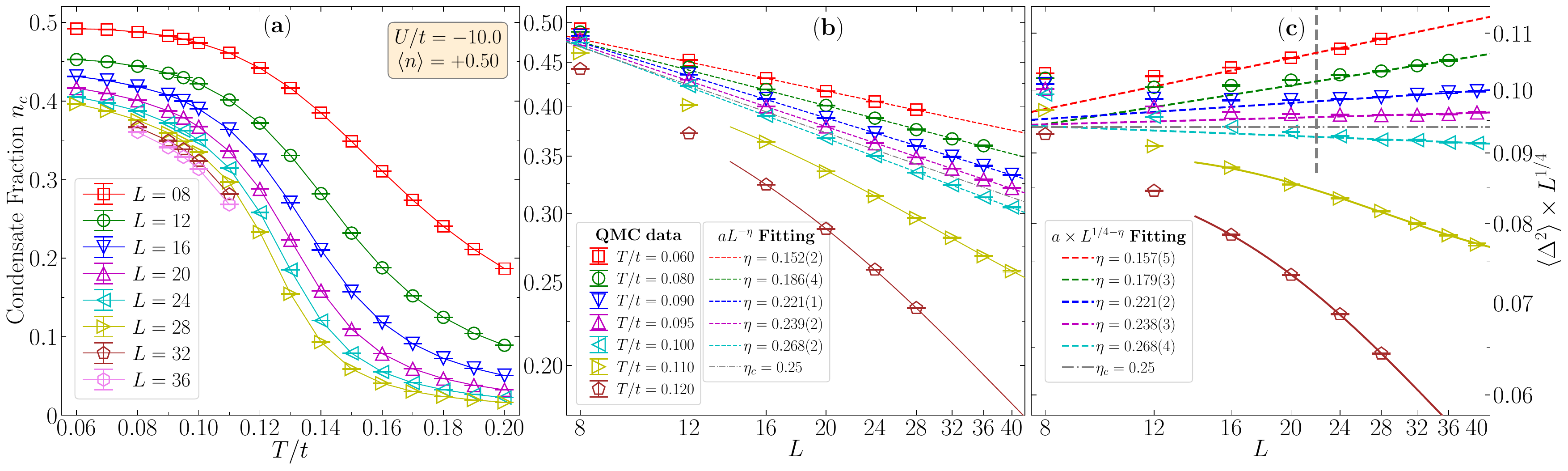}
\caption{\label{fig:Fig17CondnstU10nT050} Condensate fraction $n_c$ and on-site pairing correlator $\langle\Delta^2\rangle$ results, for $U/t=-10,\langle\hat{n}\rangle=0.50$. Panel (a) and (b) plot high-precision results of $n_c$ versus $T/t$ and $L$, respectively; (c) plots $\langle\Delta^2\rangle\times L^{1/4}$ versus $L$. Panels (b) and (c) apply log-log plot. The dashed lines (for $T/t\le0.100$) are the power-law fitting of the QMC data with $L\ge16$ in (b), and with $L\ge24$ in (c), while the solid lines (for $T/t\ge0.110$) are the exponential fitting. Consistent exponents $\eta$ are obtained from these two quantites for $T/t\le0.100$. The gray, dash-dot lines highlight the exponent $\eta_c=1/4$ for the BKT transition, with the transition temperature bounded as $0.095<T_{\rm BKT}/t<0.100$. The gray, vertical dash line in (c) marks the scaling regime as $L\ge24$ for $\langle\Delta^2\rangle$. }
\end{figure}

Figure~\ref{fig:Fig17CondnstU10nT050} plots all the results of condensate fraction $n_c$ and pairing correlator $\langle\Delta^2\rangle$ for $U/t=-10,\langle\hat{n}\rangle=0.50$. As shown in panel (b), the algebraic scaling in superfluid phase and exponential decaying in normal state of $n_c$ determine the bounded range of the transition temperature as $0.095<T_{\rm BKT}/t<0.100$. For this case, the finite-size effect in $\langle\Delta^2\rangle$ is moderate as $L\ge24$ results can reach the correct algebraic scaling and present consistent exponents $\eta(T)$ with those from $n_c$ for $T/t\le0.100$. This can be attributed to the strong attractive interaction, for which the contribution from nonlocal pairing in the system is rather small. A further scaling for the exponent $\eta(T)$ (using the $\eta$ results of $T/t=0.090$ and $0.095$) determines the transition temperature as $T_{\rm BKT}/t=0.0982(10)$.

\end{document}